\documentclass[twocolumn]{aastex61}
\usepackage{hyperref}
\usepackage{float}
\usepackage{graphicx, graphics}
\usepackage{CJK}

\usepackage{amsmath, amssymb, bm}
\newcommand*{\QED}{\hfill\ensuremath{\blacksquare}}%

\DeclareMathOperator*{\argmin}{arg\,min}

\received{}
\revised{}
\accepted{}
\published{}
\submitjournal{}

\shorttitle{NMF in Direct Imaging}
\shortauthors{R\'en et al.}

\begin{document}
\begin{CJK*}{UTF8}{gbsn}
\title{Non-negative Matrix Factorization: Robust Extraction of Extended Structures}
\author{B\=in R\'en (任彬)}\email{ren@jhu.edu}
\affiliation{Department of Physics and Astronomy, The Johns Hopkins University, Baltimore, MD 21218, USA}

\author{Laurent Pueyo}\email{pueyo@stsci.edu}
\affiliation{Space Telescope Science Institute (STScI), Baltimore, MD 21218, USA}

\author{Guangtun Ben Zhu}\email{guangtun.ben.zhu@gmail.com}
\altaffiliation{Hubble Fellow}
\affiliation{Department of Physics and Astronomy, The Johns Hopkins University, Baltimore, MD 21218, USA}

\author{John Debes}
\affiliation{Space Telescope Science Institute (STScI), Baltimore, MD 21218, USA}

\author{Gaspard Duch\^ene}
\affiliation{Astronomy Department, University of California, Berkeley, CA 94720, USA}
\affiliation{Universit\'{e} Grenoble Alpes, CNRS, IPAG, F-38000 Grenoble, France}

\begin{abstract}
We apply the vectorized Non-negative Matrix Factorization (NMF) method to post-processing of direct imaging data for exoplanetary systems such as circumstellar disks. NMF is an iterative approach, which first creates a non-orthogonal and non-negative basis of components using given reference images, then models a target with the components. The constructed model is then rescaled with a factor to compensate for the contribution from a disk. We compare NMF with existing methods (classical reference differential imaging method, and the Karhunen-Lo\`eve image projection algorithm) using synthetic circumstellar disks, and demonstrate the superiority of NMF: with no need for prior selection of references, NMF can detect fainter circumstellar disks, better preserve low order disk morphology, and does not require forward modeling. As an application to a well-known disk example, we process the archival  {\it Hubble Space Telescope} ({\it HST}) STIS coronagraphic observations of HD~181327 with different methods and compare them. NMF is able to extract some circumstellar material inside the primary ring for the first time. In the appendix, we mathematically investigate the stability of NMF components during iteration, and the linearity of NMF modeling.
\end{abstract}


\section{Introduction}
The detection and characterization of circumstellar disks and exoplanets requires very high contrasts relative to the host star. Achieving these contrasts requires a combination of telescope and instrument design to suppress the residual diffraction pattern of a star as well as state-of-the-art post-processing techniques to suppress residual systematic flux that remains. Well-designed instruments are able to stabilize the temporally varying noise in telescope exposures, creating quasi-static features \citep{perrin04, hinkley07, perrin08, soummer07, traub10}. These quasi-static features (speckles), together with the stellar point spread function (PSF), can be empirically modeled and removed by post-processing techniques, revealing circumstellar disks and exoplanets around stars \citep[e.g., ][]{marois06, lafreniere07, lafreniere09, soummer12}.

Post-processing techniques have been evolving in the past decade. When the stellar PSF and the speckles are stable and do not have significant variation over time, they can be removed by subtracting the image of a reference star (i.e., Reference-star Differential Imaging, RDI). This classical RDI method has been extensively used, unveiling revolutionary results especially for bright disks \citep[e.g., ][]{smith84, grady10, schneider14, schneider16, debes13, debes17}. When the speckles do vary, especially for ground-based AO-fed telescopes, several techniques have been proposed to resolve this (e.g., Angular Differential Imaging, ADI: \citealp{marois06}; Spectral Differential Imaging, SDI: \citealp{biller04}).

Current successful removal of both stellar PSF and quasi-static speckles are based on the utilization of innovative statistical methods \citep{mawet12}. Using a large sample of reference PSFs, there are two widely-known advanced post-processing algorithms: one is the Locally Optimized Combination of Images algorithm (LOCI, \citealp{lafreniere07}), the other is the principal component analysis based Karhunen-Lo\`eve Image Projection algorithm (KLIP, \citealp{soummer12, amaya12}). Both LOCI and KLIP have many new discoveries, including with archival data \citep[e.g.][]{lafreniere09, soummer12, soummer14, choquet14, choquet16, mazoyer14, mazoyer16}. However, the aggressive PSF subtraction of LOCI biases the photometry and astrometry of the results \citep{marois10, pueyo12}, and the projection onto the eigen-images of KLIP reduces the flux from astrophysical objects and requires forward modeling to recover it \citep{soummer12, choquet16, choquet17}. Most forward modeling techniques are designed for exoplanets \citep{pueyo16} though attempts have been made for circumstellar disks \citep[e.g., ][]{esposito14, milli14, wahhaj15, follette17}. Irregularly-shaped disks (see canonical examples in \citealp{grady13, dong16}) do not benefit from forward modeling since it is difficult to recover the initial surface brightness distribution \citep{follette17}.

An accurate recovery of disk morphology aids in the understanding of disk properties in several ways. First, one can retrieve a surface brightness profile, which can reveal possible asymmetric structures and the traces of complex dynamical structures (spiral arms, jets, clumps, etc. See \citealp{dong15, dong16} for some canonical examples). Second, one is able to study the evolution of disks on short timescales (see \citealp{debes17} for the yearly evolution of the TW Hydrae disk). Third, the morphology of disks can indicate the possible existence of unseen planets that are perturbing the circumstellar disk structure, creating observable signatures \citep[e.g.,][]{rodigas14, lee16, nesvold16, dong17}.

The aim of post-processing is to detect and characterize point sources and extended circumstellar disks. Point source detection methods have gained much progress in recent years (SDI, ADI, etc.), however all these algorithms perform poorly for extended disks, and it is difficult, if not nearly impossible, to fully calibrate the disks in that context. With an eye towards more robust and well-calibrated PSF subtraction for disk studies, we study in this paper the Non-negative Matrix Factorization (NMF) method. The early work of NMF was carried out by \citet{paatero94}, and became well-known after \citet{lee01} who provided updated rules to guarantee convergence through iteration. In \citet{blanton07} the authors determined update rules which can handle non-uniform and missing data primarily for astronomical spectroscopic observations. Their method was improved by \citet{zhu16} in a vectorized form, which is adopted in this paper because of its excellent parallel computational efficiency. An early attempt with NMF on high-contrast imaging has been performed by \citet{gonzalez17}; in this paper we study the method in detail.

In post-processing, the steps of NMF are similar to KLIP: construct components from given references first, then model any target with the components. Unlike KLIP, NMF does not remove the mean of every image, and keeps all the entries non-negative, which consequently constructs a non-orthogonal component basis. When modeling a target, non-negative coefficients for the components are obtained. In order to perfectly recover the morphology of the astrophysical signals such as circumstellar disks, the NMF subtraction results do need forward modeling, however we are able to show that, to the first order, this can be performed with a simple search in a one-dimensional space.

The structure of this paper is as follows: \textsection\ref{sec:methods} discusses the limitation of current methods and explains the mechanism of NMF;  \textsection\ref{sec:application} is composed of the post-processing results of various methods with modeled disks, and the application to a classical example of HD~181327, thus making the NMF method standing out among current ones; and \textsection\ref{sec:summary} summarizes the general performance of NMF, and discusses its significance to the field. In the appendices, Appendix~\ref{appendixSymbols} is a list of symbols used in this paper; Appendix~\ref{appendixRules} shows the update rules proposed by \citet{zhu16}, as well as the adjustments made for direct imaging data; Appendix~\ref{compoConstr} investigates the stability of NMF components during their construction, Appendix~\ref{scalefactorproof} presents how to model a target with the NMF components, and Appendix~\ref{BFFwhy} describes the our procedure to correct for over-subtraction.

\section{Methods}\label{sec:methods}
To illustrate the limitation of current methods and compare them with NMF, we took data from the {\it HST} Space Telescope Imaging Spectrograph (STIS) coronagraphic observations of HD 38393 ($\gamma$ Leporis, Proposal ID: 14426\footnote{\url{http://archive.stsci.edu/proposal_search.php?mission=hst&id=14426}}, PI: J.~Debes), aligned the centers of the stars using a Radon Transform-based center determination method described in \citet{pueyo15}. We release an improved center-determination code {\tt centerRadon}\footnote{\url{https://github.com/seawander/centerRadon}.}, which is based on Radon Transform and performs line integrals on specific parameter spaces and selected regions. The code is at least 2 orders of magnitude faster than classical Radon Transform algorithms for our STIS data.

We cut the aligned exposures into $87\times87$ pixel arrays\footnote{Note: the dimension of all the images in this paper is $87\times87$ pixels ($4.41\times4.41$ arcsec$^2$) unless otherwise specified.}, corresponding with a $4.41\times4.41$ arcsec$^{2}$ field of view. The original data contains 810 $0.2$-second exposures of HD~38393 at 9 different telescope orientations. In our simulation, all the data are normalized to have flux units of mJy arcsec$^{-2}$; and to save computational time, we use only $81$ exposures by selecting just $9$ exposures at each orientation.

This section is organized as follows: \textsection\ref{currentLimits} discusses the limitation of current modeling methods, \textsection\ref{nmf} explains the NMF method in detail.

\subsection{Limitation of Current Methods}\label{currentLimits}
Current post-processing methods are excellent at finding disks, but they can have disadvantages and particular limitations. We start our simulation with a synthetic face-on disk generated by {\tt MCFOST}, a radiative transfer software to model circumstellar disks with particular morphologies and compositions \citep{pinte06, pinte09}. For the exposures of HD~38393 at one orientation, the synthetic disk was injected after being added with Poisson noise to simulate real observations, therefore creating the {\it target} images; while the exposures from the other 8 orientations are treated as the {\it references}.

The targets and references are then used for post-processing. In order to perform the RDI technique, we compute the pixel-wise median of the references, creating an empirical model of the stellar PSF and speckles, and subtract the model from the disk-injected images. For KLIP, we construct the components from the references, and model the targets with the components, then remove the KLIP model from the targets. This procedure is then executed for all the $9$ orientations, then the final result is calculated from the pixel-wise median of the 81 subtracted targets.

\begin{figure}[htb!]
\center
\includegraphics[width=0.48\textwidth]{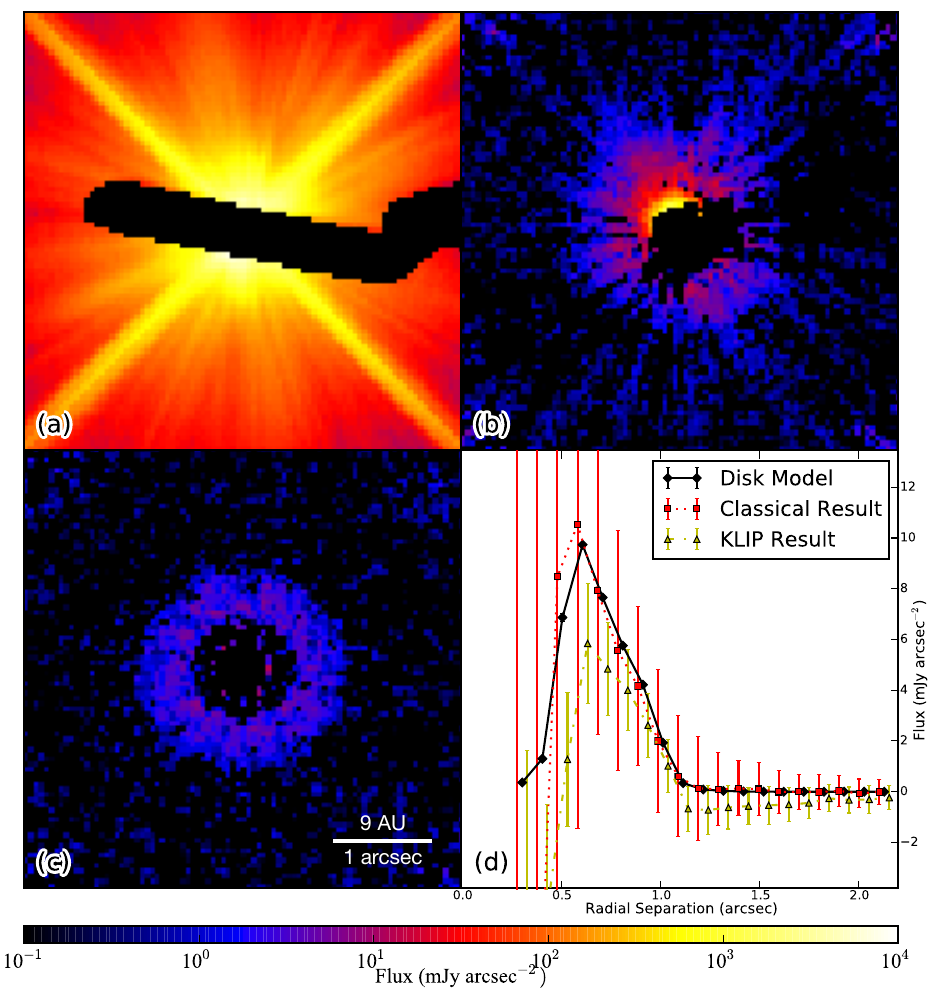}
\caption{Demonstration of limitations to current PSF subtraction methods using a synthetic face-on circumstellar disk with integrated flux ratio $F_{\mathrm{disk}}/F_{\mathrm{star}} = 7.4\times10^{-6}$. ({\bf a}) The {\it HST}-STIS exposures of HD~38393 added with a synthetic face-on disk and no PSF subtraction. The disk is unseen due to its faintness relative to the wings of the stellar PSF, and the central dark region is the coronagraphic BAR5 mask of STIS. ({\bf b}) Classical RDI subtraction result: the north-eastern region shows an over-luminosity which does not belong to the disk. ({\bf c}) Subtraction result with KLIP: the disk is seen but its flux is reduced and its morphology modified. ({\bf d}) Radial surface brightness profiles of the subtraction results and the disk model. While the profile with the classical method agrees with the disk model, the disk is not significantly detected in this region due to large systematic PSF subtraction residuals. KLIP is unable to recover the true radial profile, introducing unphysically negative pixels to the surface brightness distribution. See Fig.~\ref{fig3:compareGeometry20x} for the disk model, and the subtraction result with NMF.}
\label{fig1:currentLimits}
\end{figure}

The subtraction results with a synthetic face-on disk are shown in Fig.~\ref{fig1:currentLimits}. The classical RDI subtraction method does not significantly recover the disk. KLIP significantly recovers the general morphology of the disk, but loses the details (e.g., the slope of the radial profile). KLIP over-fits the disk and thus reduces the flux \citep{choquet14, ren17}. For the radial profiles of the recovered disks, although the azimuthally averaged flux of the classical RDI result agrees with that of the disk model, this consistency is not statistically significant -- the uncertainty for a given radial separation bin is calculated from the standard deviation of the pixels located within $\pm 0.5$ pixel\footnote{The STIS pixel size corresponds with 0.05078 arcsec \citep{stisihb17}.} from that radial location: large azimuthal variations result into large standard deviations. Although the uncertainties for the KLIP result are smaller, there is a systematic downward bias. More importantly, the slope of the KLIP result is not consistent with that of the true surface brightness profile of the model disk, and forward modeling has to be performed to reveal the true distribution.

In order to accurately retrieve the morphology and photometry of planets or disks simultaneously, forward modeling for KLIP post-processing has been adopted in the literature. However, this procedure not only assumes {\it prior} understanding of the objects \citep[e.g.,][]{soummer12, wahhaj15, pueyo16}, it is also time-consuming to iteratively recover the likely disk surface brightness distribution \citep[e.g.,][]{choquet16, choquet17}.

\subsection{Non-negative Matrix Factorization (NMF)}\label{nmf}
The methods to faithfully recover astrophysical objects with high contrast imaging techniques has been evolving. On one hand, new methods have been proposed and studied to minimize over-subtraction: \citet{pueyo12} focuses on the positive coefficients for the LOCI method, and substantially improves the characterization quality of point source spectra. On the other hand, forward modeling is introduced as a correction method for the reduced data: \citet{wahhaj15} assumes a prior model of disks for LOCI subtraction. \citet{pueyo16} takes the instrumental PSF to characterize point sources with the KLIP method. Current forward modeling attempts are best optimized for planet characterization, while for extended and resolved objects like circumstellar disks, assumptions of disk morphology have to be made. These assumptions may not accurately recover the true flux, particularly for disks that deviate from simple morphologies. In this paper, we aim to circumvent the forward modeling difficulties by studying a new method -- Non-negative Matrix Factorization (NMF).

NMF decomposes a matrix into the product of two non-negative ones \citep{paatero94, lee01}, a technique that has been used over the past decade to account for astrophysical problems \citep{blanton07, zhu16}. Inspired by the \citet{pueyo12} work of adopting positive coefficients, we study NMF because of its non-negativity, which is well suited for astrophysical direct imaging observations. The previous applications of NMF are to one-dimensional astrophysical spectra. Two-dimensional images have  significantly larger amounts of information and therefore escalate the computational cost. In order to make this problem computationally tractable we make the following adjustments: we flatten every image into a one-dimensional array to maximize the utilization of currently available tools and we adopt the vectorized NMF technique \citep[][{\tt NonnegMFPy}\footnote{\url{https://github.com/guangtunbenzhu/NonnegMFPy}}]{zhu16} to implement parallel computation with multiple cores.

The NMF application to imaging data is comprised of three steps: constructing the basis of components with the reference images (\textsection\ref{componentsbuilding}), modeling any new target with the component basis (\textsection\ref{targetmodeling}), and  correcting for the over-fitting with a scaling factor (\textsection\ref{bff}). We release our version of NMF \footnote{\url{https://github.com/seawander/nmf_imaging}}, which is also available in the {\tt pyKLIP} package \citep{pyklip15}.

\subsubsection{Component Construction}\label{componentsbuilding}
The first step of NMF is to approximate the {\it reference} matrix $R$, with the product of two non-negative matrices: the coefficient matrix $W$, and the component matrix $H$, i.e., \begin{equation}\label{eq1}
R \approx W H,
\end{equation}
by minimizing their Euclidean distances, see Appendix~\ref{appendixSymbols} for detailed definition of symbols. The approximation of Eq.~\eqref{eq1} is guaranteed to converge with iteration rules in \citet{zhu16} using:
\begin{align}
W^{(k+1)} &= W^{(k)}\circ\frac{RH^{(k)T}}{W^{(k)}H^{(k)}H^{(k)T}}\label{coefUpdate},\\
H^{(k+1)} &= H^{(k)}\circ\frac{W^{(k)T}R}{W^{(k)T}W^{(k)}H^{(k)}}\label{compUpdate},
\end{align}
with random initializations. In the above equations, the circle $\circ$ and fraction bar\footnote{Note: all the fraction bars in this paper are element-wise division of matrices unless otherwise specified} $\frac{(\cdots)}{(\cdots)}$ denote element-wise multiplication and division for matrices, the superscripts enclosed with $^{(\cdot)}$ denote iteration steps, and the superscript ${}^T$ stands for matrix transposition. For astronomical data, a weighting function $V$, which is usually the elementwise variance (i.e., the square of the uncertainties) of $R$, is applied to weigh the contribution from different pixels and take care of heteroskedastic data \citep{blanton07, zhu16}, see Appendix~\ref{appendixRules} for the adaptation for our STIS imaging data.

The connection between NMF and previous statistical methods can be illustrated using Eq.~\eqref{coefUpdate}: we can cross out the $W$ terms on the righthand side\footnote{Note: this operation is for demonstration purpose only, it is not mathematically practical.}, and get $W=\frac{RH^T}{HH^T}$, which stands for the projection of vector $R$ onto vector $H$. This expression is in essence performing least square estimation as in KLIP, where the inversion of the covariance matrix of the components is required (i.e., the inverse of $HH^T$), however the covariance matrices are often poorly conditioned for inversion. Intuitively, NMF returns a non-negative approximation of the matrix inverse through iteration.

In the KLIP method, the importance of the components is ranked based on the magnitude of their corresponding eigenvalues. For NMF we rank them by constructing the components {\it sequentially}: with $n$ components constructed we construct the $(n+1)$-th component using the $n$ previously constructed ones. In our construction we only randomize the initialization of the $(n+1)$-th component, while the first $n$ components are initialized with their previously constructed values.  See Appendix~\ref{compoConstr} for detailed expression and derivation. This construction method not only ranks the components, but also is essential for the linearity in target modeling in the next subsection.

\subsubsection{Target Modeling (``Projection'')}\label{targetmodeling}

The sequential construction of NMF components is the foundation of this paper. First, as illustrated in Appendix~\ref{compoConstr}, the components remain stable in this setup. Second, the stability of the components guarantees a linear separation of the disk signal from the stellar signals (Appendix~\ref{scalefactorproof}). Third and most importantly, the linearity of the target modeling process allows for our attempt to {\it circumvent forward modeling with a scaling factor}, as illustrated in Appendix~\ref{BFFwhy}.

With the basis of NMF components constructed sequentially, the next step is to model the targets with the components. For a flattened target $T$, we now minimize $||T - \omega H||^2$ with an iteration rule\footnote{This rule is the one dimensional case of Eq.~\eqref{coefUpdate}.}
\begin{equation}\label{targetmodelling}
\omega^{(k+1)} = \omega^{(k)}\circ\frac{TH^T}{\omega^{(k)}HH^T},
\end{equation}
where $\omega$ is the $1\times n$ coefficient matrix for the target, and $H$ is the NMF components constructed in the previous paragraph. This expression is essentially performing least square approximation as in KLIP, but the coefficients are smaller in magnitude (Appendix~\ref{scalefactorproof}). A more detailed expression taking weighting function into account is given by \citet{zhu16} and the adaptation to our STIS data is shown in Appendix~\ref{appendixRules}. When the above process converges, the NMF model of the target can be represented by 
\begin{equation}
T_{\text{NMF}} = \omega H.
\end{equation}

\subsubsection{Disk Retrieval via ``Forward Modelling''}\label{bff}
With sequentially constructed components, the target modeling procedure is able to linearly separate the circumstellar disk from the others, as illustrated in Eq.~\eqref{Dinfluence}. To first order, we have
\begin{equation}
T_{\text{NMF}} = D_{\text{NMF}} + S_{\text{NMF}},
\end{equation}
where the subscript ${}_{\text{NMF}}$ means performing the NMF modeling result for the stellar signal ($S$) or disk signal ($D$) alone. In addition, when we {\it sequentially} model the target, if the disk does not resemble any NMF component, then the first component, which explains $\sim90\%$ of the residual noise (shown in \textsection{\ref{noiseremove}}), will always dominate the modeling for the disk -- the captured morphology of the disk is just another copy of the NMF model of the stellar PSF and the speckles (shown in \textsection\ref{nmfVSklip-disk}), i.e., \begin{equation} D_{\text{NMF}}\approx \alpha S_{\text{NMF}},\end{equation} where $\alpha$ is a positive number.

The way to correct for the contribution from the disk is to introduce a scaling factor $\hat{f}$, which satisfies \begin{equation}\hat{f}T_{\text{NMF}}=S_{\text{NMF}}, \end{equation} then when we subtract the scaled NMF model of the target from the raw exposure, we will have the disk image:
\begin{equation}
T - \hat{f}T_{\text{NMF}} = T - S_{\text{NMF}} = D.
\end{equation}
Ideally, we will solve for $\hat{f}=1/(1+\alpha)$, however since $\alpha$ is not known, we have to find $\hat{f}$ empirically.

\begin{figure}[htb!]
\center
\includegraphics[width=0.45\textwidth]{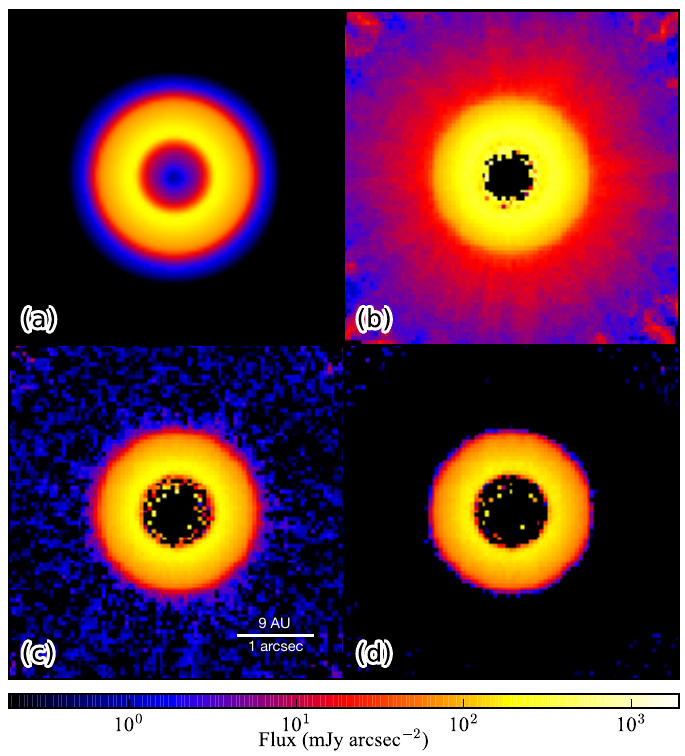}
\caption{Illustration of the scaling factor for the disk model in Fig.~\ref{fig3:compareGeometry1xKLIPvsNMF}. ({\bf a}): Face-on disk model created by {\tt MCFOST}. ({\bf b}): Scaled reduced disk with $f = 0.930$. There are PSF residuals since the scaling factor is smaller than the optimum one. ({\bf c}): Scaled reduced disk with $f = \hat{f}=0.982$, the best disk corresponding with the optimal scaling factor obtained from the BFF procedure. ({\bf d}): Reduced disk with no correction (i.e., $f = 1$). The disk flux is reduced due to oversubtraction, and pixels beyond the outskirts of the disk are all negative. See Fig.~\ref{fig5:sigmafVSdistanceB} for a comparison of the radial profiles.}
\label{fig4:sigmafVSdistanceA}
\end{figure}

\begin{figure}[htb!]
\center
\includegraphics[width=0.47\textwidth]{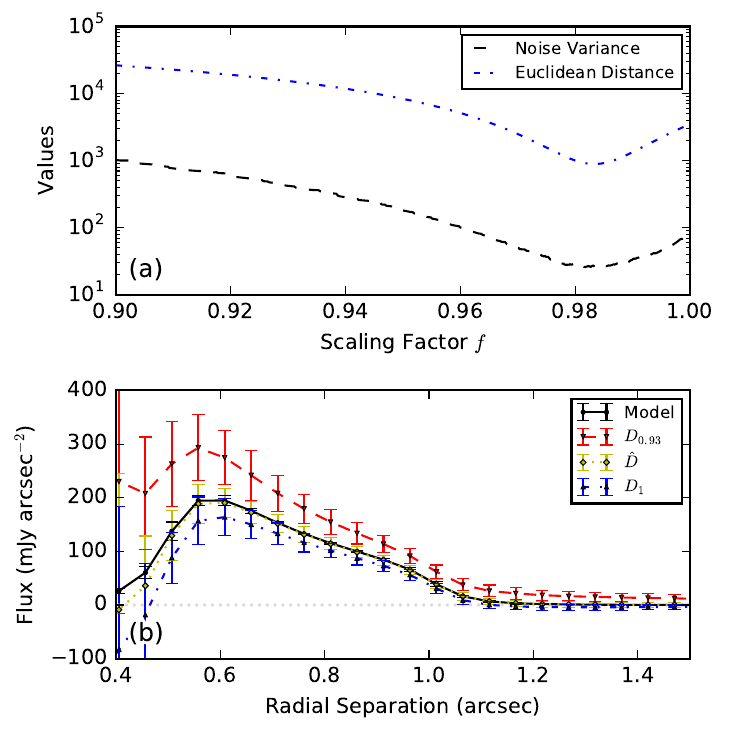}
\caption{({\bf a}) The curve for Euclidean distance between the scaled disks and the {\tt MCFOST} model in Figs.~\ref{fig3:compareGeometry1xKLIPvsNMF} and \ref{fig4:sigmafVSdistanceA} (dash-dotted blue line) is consistent with the curve for the background noise (dashed black line). This is the demonstration of the effectiveness of the BFF procedure. ({\bf b}) Radial profiles for the model (black solid line), and scaled disks with three different scaling levels. When $f = 0.930 < \hat{f}$, the radial profile (black solid line) is moving upwards in relative to that of the model; when $f = 0.982 = \hat{f}$, its radial profile (dotted yellow line) agrees with that of the model; for $f = 1 > \hat{f}$, the diskless pixels are all negative (blue dash-dotted line, compare with the gray dashed horizontal line of $0$'s). See Fig.~\ref{fig3:compareGeometry1x} for the results from other methods.}
\label{fig5:sigmafVSdistanceB}
\end{figure}

We introduce the Best Factor Finding (BFF) procedure in Appendix~\ref{BFFwhy} as our attempt to {\it circumvent forward modeling} with a simple scaling factor that minimizes the corresponding background noise. To illustrate the efficiency of BFF, we show our results in the STIS data with different scaling factors in Fig.~\ref{fig4:sigmafVSdistanceA} and Fig.~\ref{fig5:sigmafVSdistanceB}. When we do not know the existence of the astrophysical signal (i.e., the disk) {\it a priori}, the residual variance dependence on scaling factor agrees consistently with the dependence of the Euclidean distances between the reduction results ($D_f$'s) and the true model ($D$) on the scaling factor. This consistency has been observed for synthetic disks at different inclination angles in our simulation, which is not shown in this paper to avoid redundancy of figures.

We could use multiple scaling factors to rescale every over-fitting component, and BFF will work for components that are affecting the overall morphology. Given the sparseness of the NMF coefficients ($\omega_{D}$, \textsection\ref{nmfVSklip-disk}), this is easily achievable with a grid search. However, we only focus on the first component, since the BFF procedure is trying to optimize the whole field of view, while the components of higher order usually do not have much influence in our simulations.

\section{Comparison and Application}\label{sec:application}

In the previous section we have demonstrated the limitation of current methods and the mechanism of NMF. In this section, we aim to demonstrate the ability of NMF in direct imaging using specific examples: we first compare the statistical properties of NMF and KLIP, as well as the intermediate steps of them in \textsection\ref{klipvsnmf}, then focus on the post-processing results. We compare the NMF results with that of the classical RDI and KLIP subtraction methods using  synthetic disks in \textsection\ref{modelcompare}. In \textsection\ref{hd181327} we focus on a well-known example as a sanity check of NMF: applying the method to the {\it HST} -STIS coronagraphic imaging observations of HD~181327.

\subsection{NMF vs KLIP: Statistical Properties \& Intermediate Steps\label{klipvsnmf}}

\begin{figure*}[thb!]
\center
\includegraphics[width=0.9\textwidth]{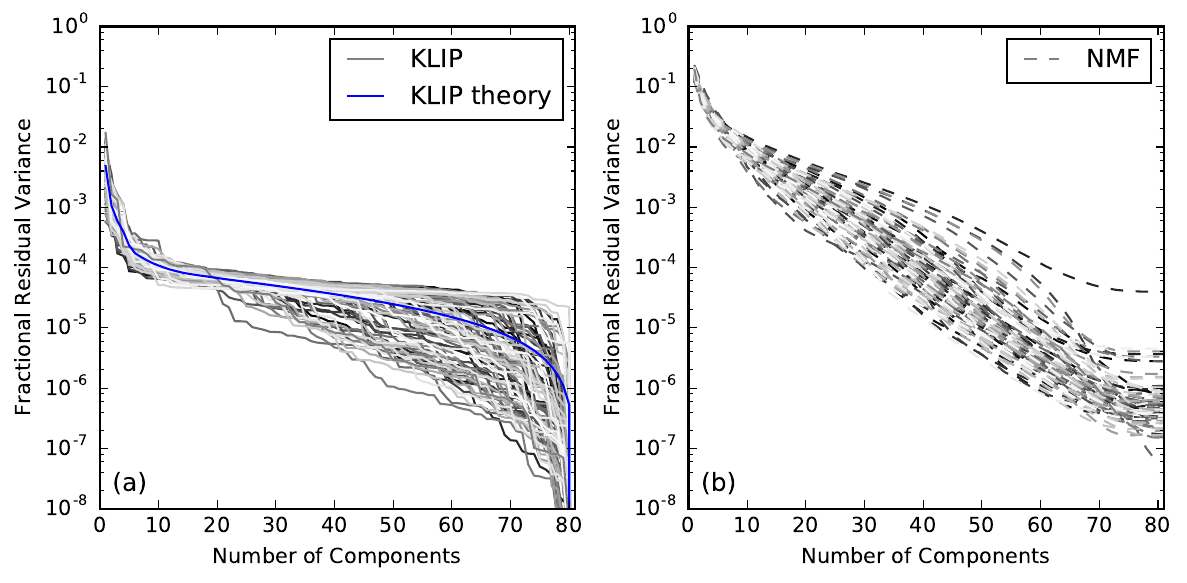}
\caption{FRV as a function of the number of components. ({\bf a}): FRV plots for KLIP, the gray solid lines are that for individual images, the blue solid line is the theoretical curve (as in Fig.~1 of \citealp{soummer12}). The existence of the plateau from $n\approx10$ to $n\approx60$ indicates that KLIP is not efficiently capturing noise over those components. ({\bf b}): FRV plots for NMF, the gray dashed lines are for individual images. There is no plateau in the NMF reduction, indicating it continues capture noise when we increase the number of components. The comparison between KLIP and NMF projections at similar FRV levels is shown in Fig.~\ref{fig3:compareGeometry1xKLIPvsNMF}. Note: The FRV trends of KLIP and NMF are not limited to the STIS data studied in this paper, and should be applicable to all other instruments.}
\label{fig2:frv}
\end{figure*}

\begin{figure*}[htb!]
\center
\includegraphics[width=0.9\textwidth]{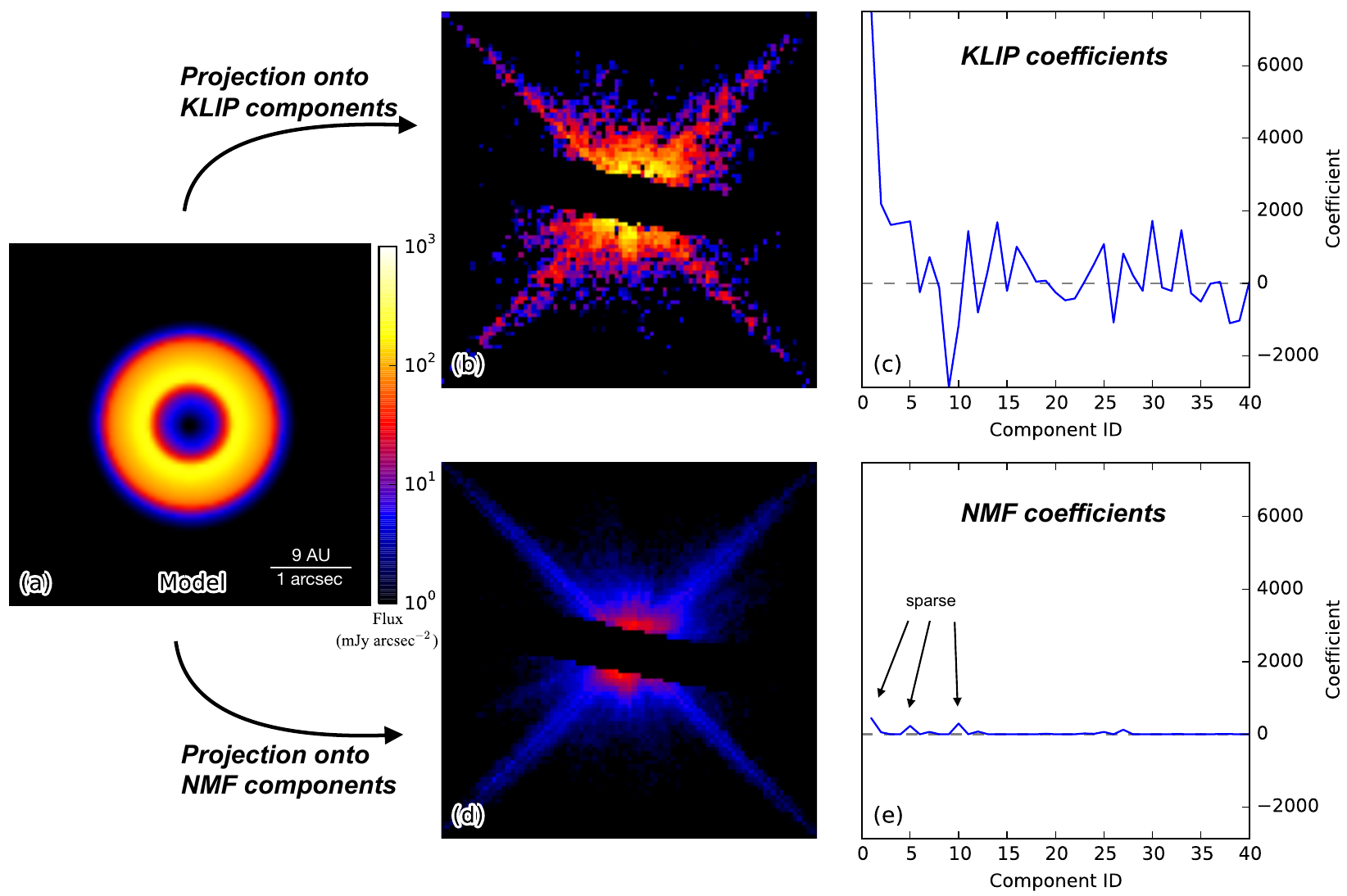}
\caption{Comparison between KLIP and NMF projections using a synthetic face-on {\tt MCFOST} disk model. ({\bf a}) The disk model. ({\bf b}) The projection of the disk model onto the KLIP components, the  central circularly-shaped structure is the result from over-fitting. ({\bf c}) The coefficients of each component in KLIP modeling. ({\bf d}) The ``projection'' of the model onto the NMF components. ({\bf e}) The coefficients of each component in NMF modeling in (d): the fact that both the components and the coefficients are non-negative reduces the likelihood of over-fitting, as shown in Eq.~\eqref{ineq}. Note: the central dark regions in (b) and (d) are the coronagraphic occulting mask at the STIS BAR5 position; and the images are in the same scale.}
\label{fig3:compareGeometry1xKLIPvsNMF}
\end{figure*}

In this subsection, we aim to address the statistical differences and intermediate steps between NMF and KLIP, and investigate why the non-negativity of NMF can yield better results. Noise and disk signal are the two constituents in a target image. However, they are always correlated with each other and separating them is the goal of all post-processing efforts. In the target modeling process, we aim to maximize noise removal and minimize disk flux removal. We therefore compare NMF with KLIP in these two aspects.

\subsubsection{Noise Removal}\label{noiseremove}

Removing the quasi-static noise from the observations is the most fundamental procedure in post-processing. With the $81$ STIS images of HD~38393, we calculate the Fractional Residual Variance (FRV) curves in the following way (Fig.~1 of \citealp{soummer12}). For each image we cumulatively increase the number of components and model it with KLIP or NMF, then subtract the model from the image to obtain the residual image. Finally the FRV is calculated by dividing the variance of the residual image by that of the original image. The comparison is shown in Fig.~\ref{fig2:frv}.

The FRV curves for KLIP decrease very fast at first, indicating KLIP is removing quasi-static noise. The curve then plateaus over many components and drops again when almost all the components are used. The existence of the plateau is when KLIP is not removing the {\it noise}, and it might even be trying to capture the {\it disk signal} if anything is fitted during the plateau. When the curves drop again, KLIP is removing the {\it{random noise}} that should not be removed using any method.

The FRV curves for NMF decrease relatively slowly while it gradually captures quasi-static noise. Eventually, NMF converges to a lower fractional residual variance than KLIP when all the components are used. \added{The higher noise level at increasing numbers of components indicates that the random noise is kept, and NMF is} preserving the difference between the target and the component basis.

\subsubsection{Disk Signal Capture}\label{nmfVSklip-disk}

Disk signal is coupled with the stellar PSF and the quasi-static speckles and some fraction of it is likely to be removed in post-processing due to over-fitting. The erosion of disk signal is why current post-processing methods need forward modeling to compensate for that over-subtraction. To measure disk signal capture, we assume the stellar PSF and speckles are perfectly removed, then project a synthetic face-on {\tt MCFOST} disk onto the components to study the target modeling process: the less disk signal that is captured the better, since this minimizes over-fitting. 

The meaning of projection is different for the two methods: for KLIP, the projection process is directly performing the dot product between the target and the components; while for NMF its ``projection'' is an iterative approach, which finds a non-negative combination of the NMF components to model the target as in Eq.~\eqref{targetmodelling}. In this paper, we do {\it not} distinguish the two processes in words, but they are {\it not} identical with respect to the different methods. 

We compare their intermediate modeling step of the disk in Fig.~\ref{fig3:compareGeometry1xKLIPvsNMF}  under similar residual noise levels ($n = 40$ components when $FRV\approx10^{-4}$, and with normalized KLIP and NMF components in Fig.~\ref{fig2:frv}):

KLIP is an ``efficient'' disk capturing method, and therefore it gives rise to over-subtraction, which requires forward modeling to compensate for. Although the morphology of the disk does not resemble any KLIP component, the disk signal is captured as a result from direct linear projection. This is the evidence of KLIP falling into the regime of over-fitting: a fraction of the disk is classified as the stellar PSF or speckle noise, then it is removed from the target image. 

NMF is ``inefficient'' in disk signal capture, causing less over-subtraction and is thus preferred in post-processing. Although the NMF target modeling process is {\it in essence} performing linear projection, the projection coefficients are sparse and have smaller magnitudes than direct projections. As is shown in Eq.~\eqref{ineq}, NMF modeling does not over-fit as much as direct projection methods like KLIP.

The sparsity of these coefficients inspires the ``forward modeling'' for NMF in \textsection\ref{bff}: we are able to accomplish this by rescaling the NMF model of a target with a simple factor, which is obtained from the BFF procedure as demonstrated in Appendix~\ref{BFFwhy}.

\subsection{Synthetic Disk Models}\label{modelcompare}

To compare NMF with current post-processing methods, we first synthesized three circumstellar disks with {\tt MCFOST} at different inclination angles and brightness levels. In this paper, we do not aim to fit any physical parameters of the disks as in the previous {\tt MCFOST} applications. However, the disks are still physically motivated and comprised of silicates with grain size ranging from 0.2 $\mu$m to 2,000 $\mu$m with a size distribution power-law index of 3.5. Their morphology are rings with a flaring index of $1.125$ spanning from $0.5$ arcsec to $1.0$ arcsec. Our {\tt MCFOST} disk models are synthesized at 0.6 $\mu$m, and convolved with the STIS TinyTim PSF \citep{krist11}{\footnote{\url{http://www.stsci.edu/hst/observatory/focus/TinyTim}}}, to simulate the {\it HST}-STIS response: at this wavelength, the incident photons from the host star are scattered by the disks and then received by the telescopes. 

\begin{figure}[htb!]
\center
\includegraphics[width=0.5\textwidth]{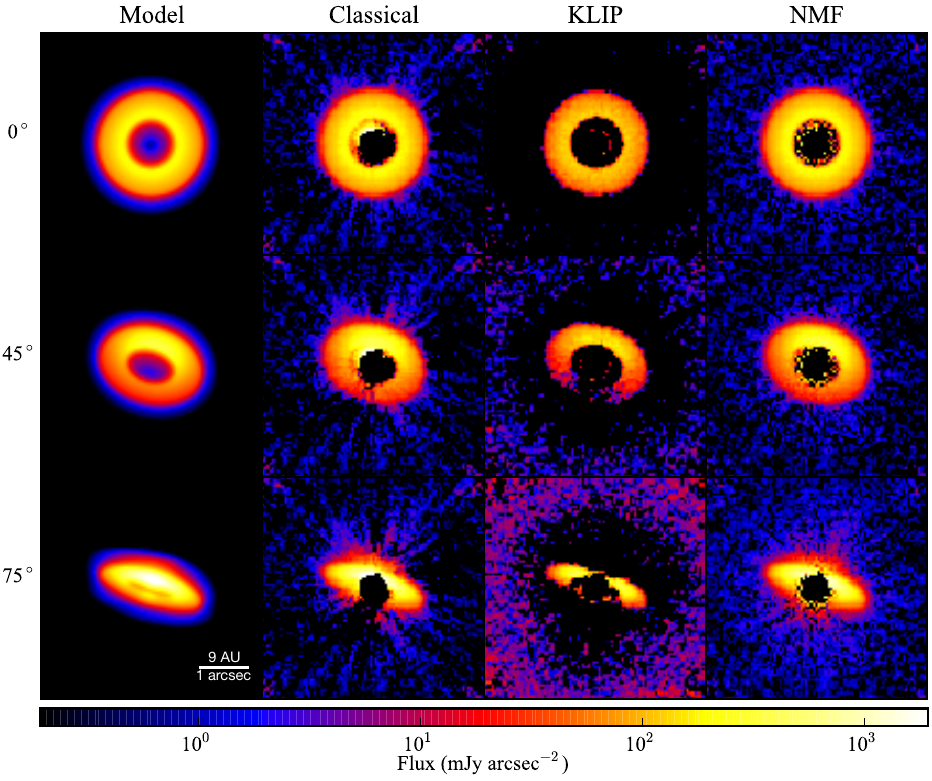}
\caption{Initial models created by {\tt MCFOST} at three different inclinations: morphology of disks reduced by different methods. From top to bottom, the disks are inclined by $0^\circ, 45^\circ,$ and $75^\circ$ (going from face-on to nearly edge-on) with $F_{\mathrm{disk}}/F_{\mathrm{star}} = (1.5, 0.9, 1.9) \times 10^{-4}$, respectively. 1st column: models;  2nd column: classical subtraction results; 3rd column: KLIP subtraction results; 4th column: NMF subtraction results. Both KLIP and NMF recover the geometries better than the classical method, and the dark halo around the KLIP images arises from its over-subtraction.}
\label{fig3:compareGeometry1x}
\end{figure}

\begin{figure}[htb!]
\center
\includegraphics[width=0.5\textwidth]{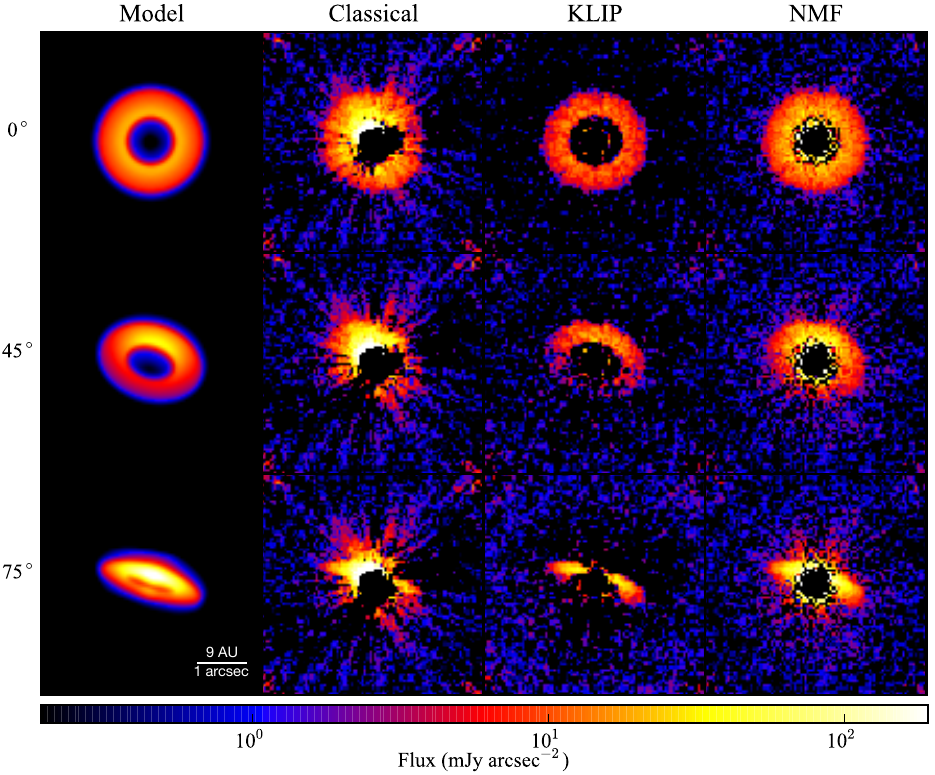}
\caption{Initial models dimmed by a factor of $10$: morphology of disks reduced by different methods for different inclination angles with $F_{\mathrm{disk}}/F_{\mathrm{star}} = (1.5, 0.9, 1.9) \times 10^{-5}$ from top to bottom. The classical method is working poorly, and NMF works better than KLIP in the sense of recovering faint signals (i.e., the far side of the inclined disks).}
\label{fig3:compareGeometry10x}
\end{figure}

\begin{figure}[htb!]
\center
\includegraphics[width=0.5\textwidth]{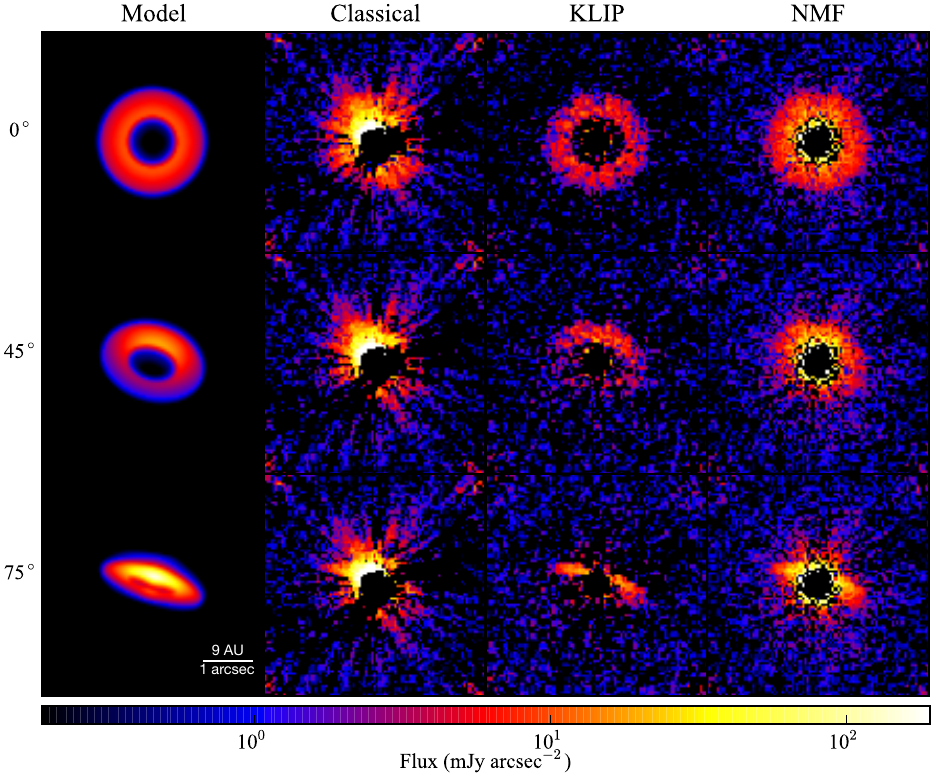}
\caption{Initial models dimmed by a factor of $20$: morphology of disks reduced by different methods for different inclination angles with $F_{\mathrm{disk}}/F_{\mathrm{star}} = (7.4, 4.8, 9.0) \times 10^{-6}$, respectively. Classical method is not working. Both KLIP and NMF recover the geometries, however NMF preserves the morphology and flux better than KLIP.}
\label{fig3:compareGeometry20x}
\end{figure}

\begin{figure}[htb!]
\center
\includegraphics[width=0.5\textwidth]{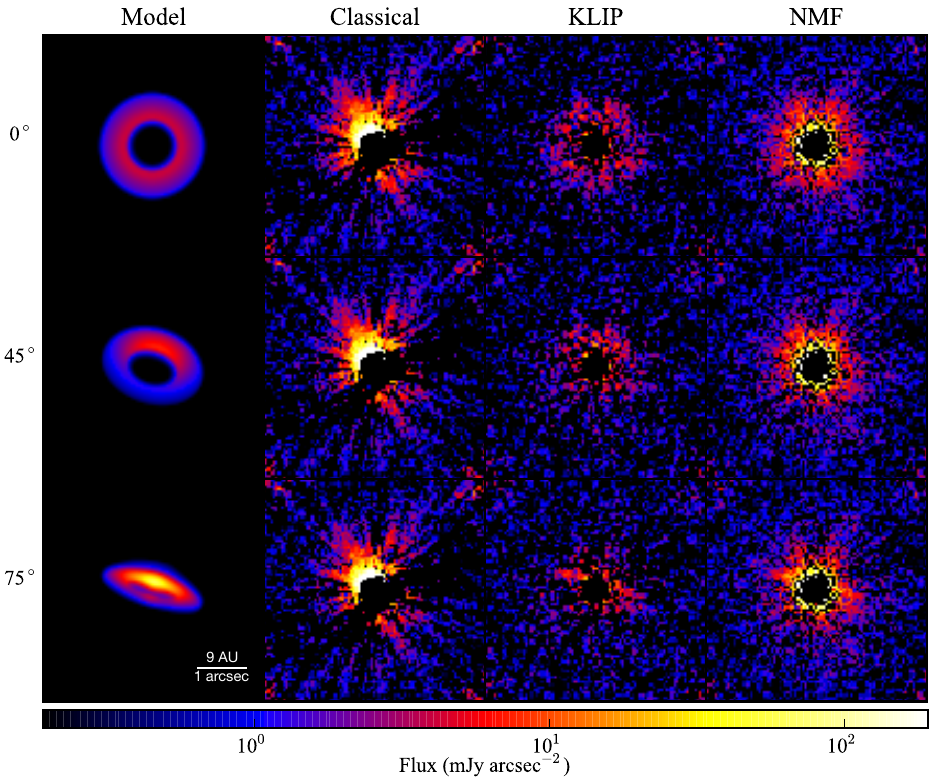}
\caption{Initial models dimmed by a factor of $50$: morphology of disks reduced by different methods for different inclination angles with $F_{\mathrm{disk}}/F_{\mathrm{star}} = (3.1, 1.9, 3.7) \times 10^{-6}$. The disks are too faint in this case, none of the methods could recover the flux of the disks properly, but NMF is still able to marginally recover the morphology.}
\label{fig3:compareGeometry50x}
\end{figure}

We reduced the synthetic disks in the same manner as in \textsection{\ref{currentLimits}} with the classical RDI, KLIP, and NMF methods. The face-on disk spanning from 0.5 arcsec to 1.0 arcsec in Fig.~\ref{fig3:compareGeometry1xKLIPvsNMF} is adopted as the initial model. We then inclined the model disk at $45^\circ$ and $75^\circ$ to verify performance for differing inclination angles. To investigate the performance of the three methods at different contrasts Fig.~\ref{fig3:compareGeometry1x} shows comparisons for disks divided by factors of $10$, $20$, and $50$, which is equivalent to reducing disk mass. The disks' $F_{\text{disk}}/F_{\text{star}}$ range from $\sim 10^{-4}$ to $\sim 10^{-6}$. The morphology results are shown in Figs.~\ref{fig3:compareGeometry1x}, \ref{fig3:compareGeometry10x}, \ref{fig3:compareGeometry20x}, and \ref{fig3:compareGeometry50x}, respectively.  

\subsubsection{Morphology}

Morphology probes the spatial distribution of the circumstellar material, which should be recovered as close to the photon noise limit as possible. From the morphology results for different disks and brightness levels in Figs.~\ref{fig3:compareGeometry1x} -- \ref{fig3:compareGeometry50x}, we compare the three methods as follows:

Classical subtraction is only able to recover the morphology of the face-on disk in the brightest cases ($F_{\mathrm{disk}}/F_{\mathrm{star}} \sim10^{-4}$), it cannot recover the dimmer ones.

KLIP is able to recover the general morphology of the disks in all cases, however the disk fluxes are reduced due to its systematic over-fitting bias (\textsection\ref{nmfVSklip-disk}), therefore forward modeling is needed to recover the real morphology of the disks. Moreover, for detailed structures, especially when the dynamic range is large, KLIP will lose the relatively faint information due to mean-subtraction (see the disappearance of the far side in the lower regions of the inclined disks in Figs.~\ref{fig3:compareGeometry1x} -- \ref{fig3:compareGeometry50x}).

NMF outperforms other methods not only in recovering the morphology of the disks, but also in recovering the faint structure: the faint far side of the inclined disks is recovered in the results. 

When the disks are too faint (Fig.~\ref{fig3:compareGeometry50x}), the reduced results are dominated by random noise, and none of the methods but NMF could marginally recover the morphology of the disks.\\

\begin{figure}[htb!]
\center
\includegraphics[width=0.45\textwidth]{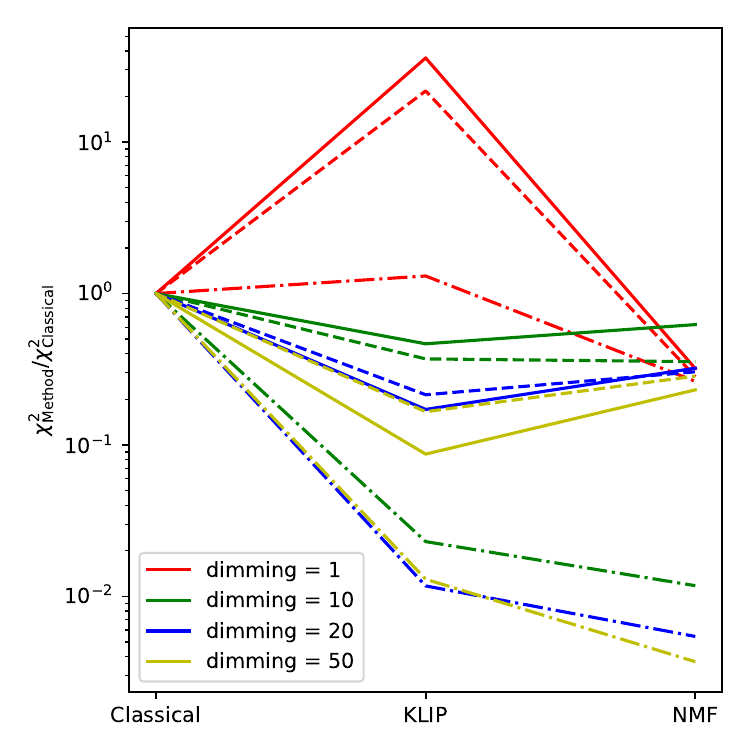}
\caption{The $\chi^2_{\text{Method}}/\chi^2_{\text{Classical}}$ ratios for different methods and dimming levels in Figs.~\ref{fig3:compareGeometry1x} -- \ref{fig3:compareGeometry50x}. The solid lines are for the face-on disk, the dashed lines are for the disks tilted at $45^\circ$, and the the dash-dotted lines are for the disks tilted at $75^\circ$. NMF is able to perform better than the classical method in the $\chi^2$ sense; in comparison with KLIP, NMF is able to reach lower or similar levels of $\chi^2$ values.}
\label{fig:chi2}
\end{figure}

For a quantitative comparison of the three methods, we compute the $\chi^2$ values using the initial model and the results in this subsection. To better illustrate the relative goodness of recovery, we plot the $\chi^2$ ratios between different methods and the classical RDI method in Fig.~\ref{fig:chi2}. NMF performs better than the classical method in this way; when comparing with KLIP, NMF is able reach lower or similar levels, demonstrating its competence in disk retrieval. In the cases when the NMF $\chi^2$ values are slightly larger than that of KLIP, it is from the fact that KLIP is over-fitting the random noise, which in principle should not be fitted by any method, rather than KLIP has a better matching to the disk model.

\subsubsection{Radial Profile}
For a face-on disk, its radial profile informs us of the spatial distribution of the amount of circumstellar material, which should be recovered faithfully. From the radial profiles shown in Fig.~\ref{fig4:compareRadial20x} for the recovered face-on disks in Fig.~\ref{fig3:compareGeometry10x}, we compare the three methods as follows:

\begin{figure}[hbt!]
\center
\includegraphics[width=0.47\textwidth]{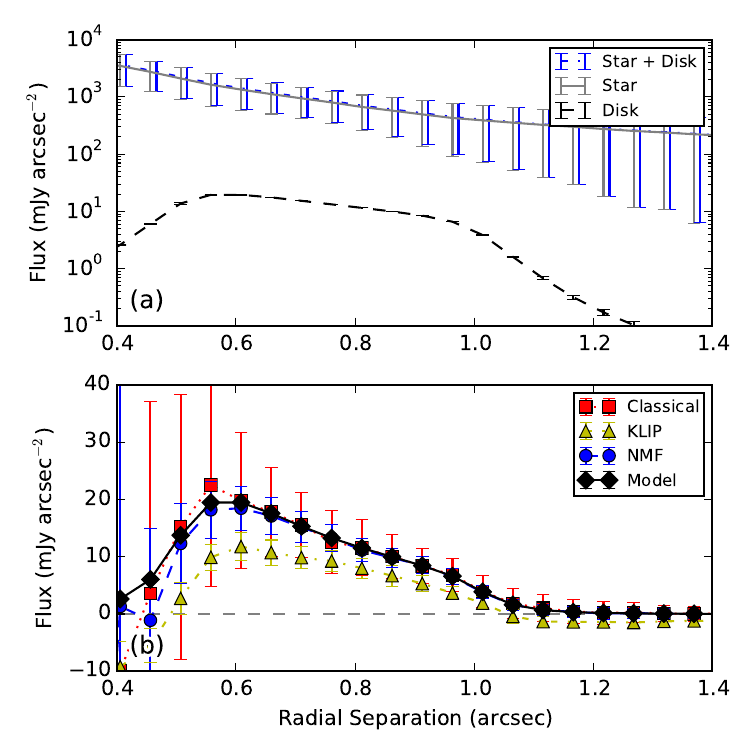}
\caption{Radial profiles for the face-on disks in Fig.~\ref{fig3:compareGeometry10x}. ({\bf a}): Radial profiles of the star, the face-on {\tt MCFOST} disk model, and the target (star added with disk). Blue dash-dotted line: the target; gray solid line: the star; black dashed line: the disk. The face-on disk is $\sim 100$  times fainter than the stellar PSF wing. ({\bf b}): Radial profiles of the disk reduced with different methods. Black diamond with solid line: disk model; red square with dotted line: classical subtraction; yellow triangle with dash-dotted line: KLIP result; blue circle with dashed line: NMF result. KLIP is over-fitting therefore introducing unphysical negative pixels in the outskirts (radial separation of more than $1.0$ arcsec), but the radial profiles of classical and NMF results are both consistent with the model, but NMF performs better with smaller uncertainties.}
\label{fig4:compareRadial20x}
\end{figure}

Classical subtraction seems to be able to recover the radial profile of the face-on disk at first glance, but it has large uncertainties. This is because we calculate the uncertainties from the standard deviation of pixels at similar radial separations. Because classical subtraction cannot suppress quasi-static noise, a radial profile with large uncertainty is typically not useful for further analysis beyond a marginal detection.

KLIP is not able to recover the radial profile of the face-on disk. This results from the over-fitting of the astrophysical signals (as discussed in \textsection\ref{klipvsnmf}): KLIP is not only unable to recover the flux correctly, it is also changing the slope of the radial profile, and forward modeling has to be implemented to recover the distribution. Although the uncertainties of KLIP are smaller, this is a result from an artificial over-fitting of the noise and does not encapsulate systematic uncertainties.

NMF not only recovers the radial profile with no bias, it also has small uncertainties. With small uncertainties, NMF is expected to detect fainter structures than the other two methods, especially for low inclination. Therefore, the NMF results can be used to perform more detailed analysis with fewer underlying assumptions, e.g., \citet{stark14}.\\

In this subsection, we have demonstrated that NMF outperforms current methods with synthetic circumstellar disks both in morphology and in radial profile. In the next subsection, we will apply NMF to a specific case when the classical method works, ensuring the reliability of NMF using a well-characterized disk.

\subsection{Application to HST-STIS Observations: HD~181327}\label{hd181327}
Unseen planets are able to perturb the circumstellar disk structure and create observable signatures \citep[e.g.,][]{jangcondell07, jangcondell12, jangcondell13, dong15, dong15a, zhu15}, and faithful recovery of both the morphology and radial profile of circumstellar disks is able to constrain the mass of these hypothetical planets \citep[e.g.,][]{rodigas14, nesvold16, dong17}. In this subsection, we aim at checking the effectiveness of NMF with a known circumstellar disk surrounding HD~181327, which ensures the reliability of NMF by comparing to a well-characterized disk.

\begin{figure*}[htb!]
\center
\includegraphics[width=0.97\textwidth]{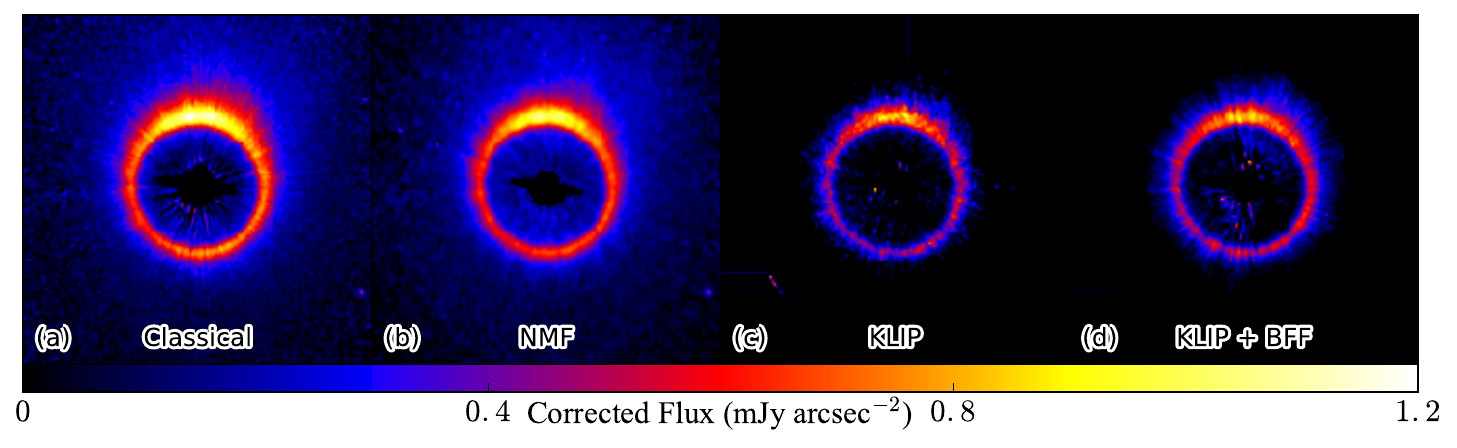}
\includegraphics[width=0.34\textwidth]{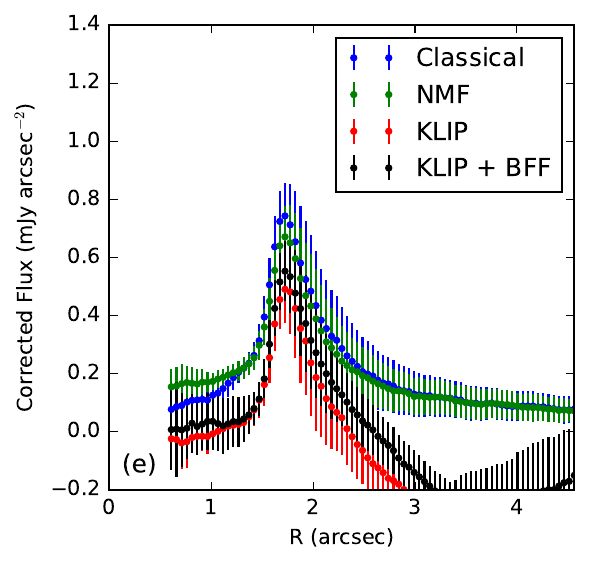}
\includegraphics[width=0.6\textwidth]{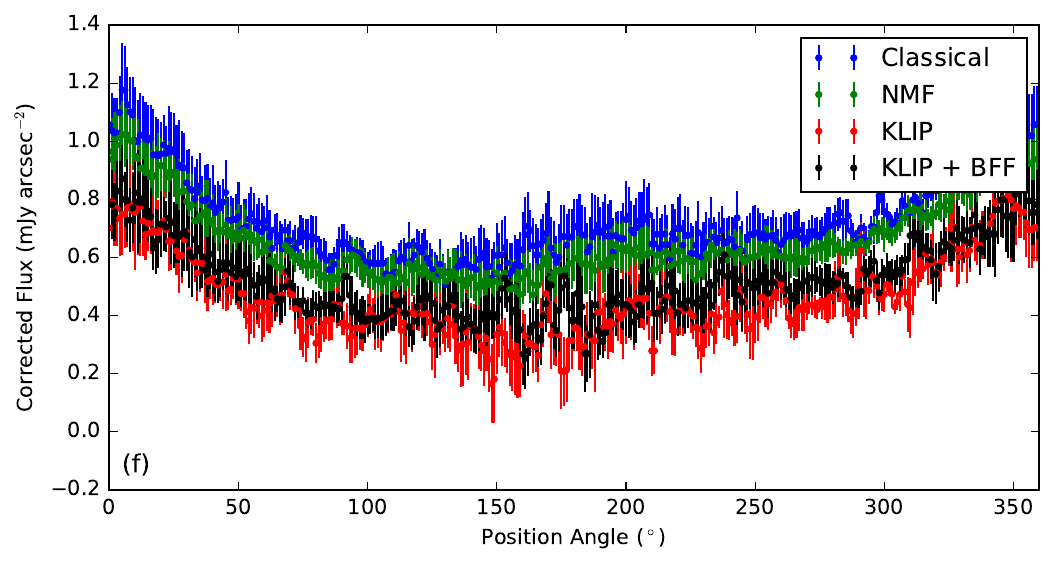}
\includegraphics[width=0.98\textwidth]{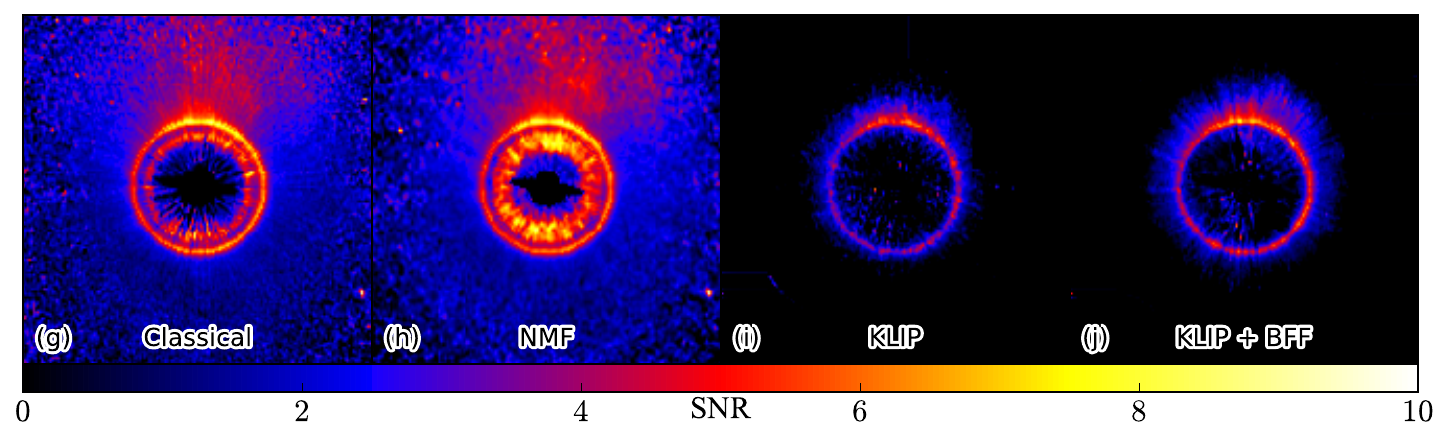}
\caption{Comparison of HD~181327 STIS disk reduced with classical subtraction, NMF, and KLIP (image dimension: $181\times181$ pixels, $9.18''\times9.18''$). ({\bf a}) Classical subtraction result of \citet{schneider14}, de-projected and illumination-corrected (i.e., $r^2$-corrected) as a pseudo face-on disk using the ellipse parameters as in \citet{stark14}. ({\bf b}) NMF subtraction result, corrected in the same way as in (a). ({\bf c}) KLIP subtraction result, corrected in the same way as in (a) and (b). ({\bf d}) KLIP subtraction result, corrected with BFF, then in the same way as in (a) and (b). ({\bf e}) Radial profiles for (a), (b), (c), and (d). ({\bf f}) Azimuthal profiles at the peak of the ring, parameters taken from \citet{stark14}. ({\bf g}), ({\bf h}), ({\bf i}), and ({\bf j}) are the SNR maps for (a), (b), (c) and (d), respectively -- in the close-in regions (inside the primary ring), NMF is able to reach higher SNR than the other methods. The results in (e) and (f) show that the NMF and classical results are mainly consistent within $1\sigma$. The KLIP results are systematically fainter due to the over-fitting of KLIP; even when corrected with the BFF procedure, the KLIP result is not convincing in either aspect.}
\label{hd181327figs}
\end{figure*}

We obtain all public {\it HST}-STIS coronagraphic observations available in December 2016 from the {\it HST} archive\footnote{\url{http://archive.stsci.edu/hst/search.php}}, and focus on the Wedge A0.6 position, then align the images with {\tt centerRadon}, and classify the exposures into two categories as in \citet{ren17}: {\it targets} which have infrared (IR) excess in their spectral energy distributions \citep{chen14}, where the IR excess is expected to emit from the circumstellar disks; and  {\it references} which do not have IR excess. After constructing the NMF components using the references as described in \textsection\ref{componentsbuilding}, we center on the observations of HD~181327 (Proposal ID: 12228\footnote{\url{https://archive.stsci.edu/proposal_search.php?mission=hst&id=12228}}, PI: G.~Schneider), which is located at $51.8$ pc and is known to host a relatively bright disk with $F_{\text{disk}}/F_{\text{star}}=1.7\times10^{-3}$ \citep{schneider14}, the disk is comprised of a nearly face-on primary ring and a faint asymmetric debris structure to the north-west \citep{stark14}.

To compare NMF with classical RDI and KLIP subtractions, we obtain the classical reduction result from \citet{schneider14}, the KLIP result with the $10\%$ closest-matching references in an $L^2$ sense \citep{ren17}, and the NMF result using the whole reference cube. We de-project and correct for the distance dependent illumination factor as in \citet{stark14}, and show the results in Fig.~\ref{hd181327figs}: the primary disk is clearly seen in all results, while the KLIP disk is systematically dimmer than the other two methods, which is the result from the over-fitting of the disk.

In terms of the morphology of the HD~181327 disk, the faint debris at the north-west region is only revealed in the classical and NMF results. Although KLIP is able to extract the primary ring, the azimuthally asymmetric outer portion of the disk is buried in the unphysical negative regions. This asymmetric debris structure was likely caused by a recent catastrophic destruction of an object with mass greater than $0.01M_{\mathrm{Pluto}}$ as determined in \citet{stark14}. With NMF, we are able to faithfully recover this structure. In the situation of fainter disks, NMF is the only method to extract the disks and retain their morphology.

Focusing on the azimuthal profile for the HD~181327 disk at the peak radial position, the majority of the NMF reduction agrees with classical subtraction both in the absolute surface brightness and in the variation with position angle to within $1\sigma$, indicating that the NMF result is capable of being studied in the classical way as in \citet{stark14}. KLIP is not consistent with the other results in either aspect. We notice that the NMF result is slightly dimmer than the classical result for the HD~181327 primary ring, which might be caused by 1) the classical subtraction may not be absolutely correct due to uncertainties in the flux scaling of reference PSFs; and 2) the BFF procedure needs diskless pixels to find the optimum scaling factor. The faint extended debris halo around HD~181327, especially the northwest debris, might be biasing the BFF scaling. However, we do not aim to argue which result better represents the disk signal, since the major purpose of this subsection is to demonstrate the excellence of NMF on well-characterized bright disks.

Another difference appears inside the primary ring of the disk. With a large number of references, NMF is able to better model the region near the inner working angle -- for the region inside the primary ring in Fig.~\ref{hd181327figs}, the NMF result has greater SNR than the classical one: the region inside the primary ring is non-zero at significance levels better than $1\sigma$, which calls for the possibility of some scattered light. The scattering dust might originate from the primary ring, which is then dragged inward by the gas in the system \citep{marino16}, or by the Poynting-Robertson drag from the radiation of the host star.

\section{Summary}\label{sec:summary}
In the post-processing of high contrast direct imaging data, the most important step is to find the best template of the stellar PSF and the speckles for a target image. Especially for broadband imaging instruments such as {\it HST}-STIS, due to the response of its filter, an ideal template to the target is a reference star with an identical spectral type. However, even if there is an ideal match, the quasi-static noise caused either by the adaptive optics system or telescope breathing will change the PSF of both the target and the reference. To capture this quasi-static noise, multiple statistical methods have been proposed and they are working most efficiently for unresolved point sources. Current advanced post-processing methods do excel in finding circumstellar disks, but their disadvantages prevent us from studying the detailed morphology of these systems.

To extract disk signals with reference differential imaging, we have demonstrated that NMF is an excellent method in capturing the stellar PSF and speckle noise. In this paper, we first compared NMF with current methods using synthetic faint disks, and demonstrated that NMF supersedes current methods both in retrieving disk morphology and in photometry; we then applied NMF to a bright disk whose morphology is well studied with the classical subtraction method, ensuring NMF is working in the most classical examples.

We propose to use NMF to overcome the limitations of current post-processing methods in extracting signals from circumstellar disks, especially to minimize over-subtraction, thus circumventing the tedious forward modeling attempts. We summarize the properties of NMF as follows:

NMF does not need reference selection to detect circumstellar disks\footnote{We still need the images of reference stars. For further detailed analyses, reference selection is preferred to get better results; otherwise the spectral types of stars should be evenly sampled.}. For broadband imaging instruments, as long as a reference library with all spectral types is given, NMF will construct the components, then find the best combination of components to model the targets. The NMF component basis can be constructed only once but works for all different targets, unless new references are added. This will be the dominant advantage of NMF in current and future big-data astronomy, e.g., surveying telescopes such as the {\it Wide Field Infrared Survey Telescope} ({\it WFIRST}). This iterative approach will need more computational time\footnote{For 72 images of dimension $87\times87$ pixels, the component construction time of NMF is $\sim 0.3 n$ minutes using 4 cores of Intel Xeon E5-2680v3 (2.5 GHz), where $n$ is the number of components that are constructed sequentially; while KLIP process takes less than 5 seconds to construct all the components with one core.}, but the gain is excellent as demonstrated in this paper.

NMF can extract disk signals and retain their morphology. The utilization of NMF will enable the study of the detailed structures and morphology of circumstellar disks especially for faint disks. With well-constrained disk morphology, we will be able to better study the formation, evolution, and even the planet-disk interaction of planetary systems.\\

With NMF, we can accomplish two goals for extracting circumstellar disks through post-processing of imaging data in this paper: detecting faint signals scattered from the disks, and recovering the morphology of them. Although our paper utilizes space-based coronagraphic observations for their excellent imaging stability, NMF is capable of capturing the varying stellar PSF and speckles from ground-based exposures (Ren et al., submitted), opening up a new way to better characterize circumstellar disks.

\acknowledgments
The authors thank the anonymous referee for the useful suggestions and comments. This work is based on observations made with the NASA/ESA Hubble Space Telescope, and obtained from the Hubble Legacy Archive, which is a collaboration between the Space Telescope Science Institute (STScI/NASA), the Space Telescope European Coordinating Facility (ST-ECF/ESA) and the Canadian Astronomy Data Centre (CADC/NRC/CSA). B.R.~thanks the useful discussions with Christopher Stark, Cheng Zhang, and Jason Wang; the comments and suggestions from Fran\c{c}ois M\'enard and Christophe Pinte; the classical subtraction result of HD~181327 provided by Glenn Schneider; and computational resources from the support of Colin Norman and the Maryland Advanced Research Computing Center (MARCC). MARCC is funded by a State of Maryland grant to Johns Hopkins University through the Institute for Data Intensive Engineering and Science (IDIES). G.B.Z.~acknowledges support provided by NASA through Hubble Fellowship grant \#HST-HF2-51351 awarded by the Space Telescope Science Institute, which is operated by the Association of Universities for Research in Astronomy, Inc., for NASA, under contract NAS 5-26555. G.D.~acknowledges funding from the European Commission's seventh Framework Program 
(contract PERG06-GA-2009-256513) and from Agence Nationale pour la Recherche (ANR) of France under contract ANR-2010-JCJC-0504-01.

\software{\\
{\tt centerRadon}: Center determination code for stellar images; {\tt MCFOST}: Radiative transfer code for circumstellar disk modeling; {\tt NonnegMFPy}: Vectorized non-negative matrix factorization code; {\tt nmf\_imaging}: Application of {\tt NonnegMFPy} on high contrast imaging.}

\facility{{\it HST} (STIS)}
\appendix
\newpage
\section{List of Symbols}\label{appendixSymbols}
\begin{deluxetable}{c c c p{12cm}}[htb!]
\tablecolumns{4} 
\tablewidth{0pc} 
\tablecaption{List of Symbols}
\tablehead{ 
\colhead{Symbol}    & \colhead{Expression}  & \colhead{Dimension} & \colhead{Meaning}}
\startdata 
$\circ$ & $(A\circ B)_{ij} = A_{ij}B_{ij}$ & & Element-wise (Hadamard) multiplication for matrices $A$ and $B$ of same dimension.\\
$D$ & & $1\times N_{\text{pix}}$ & Flattened image of the astrophysical signal (i.e., no stellar information).\\
$\hat{D}$ & $T - \hat{f}T_{\text{NMF}}$& $1\times N_{\text{pix}}$ & Reduced best image of the astrophysical signal ($D$), obtained from BFF procedure.\\
$D_f$ & $T - fT_{\text{NMF}}$& $1\times N_{\text{pix}}$ & Reduced image of the astrophysical signal with scaling factor $f$.\\
$D_{\text{NMF}}$& $\omega^{({D})} H$& $1\times N_{\text{pix}}$ & NMF model of the astrophysical signal ($D$).\\
$\delta(\cdot)$ & &  & The change of the $(\cdot)$ item after one iteration.\\
$F_{\mathrm{disk}}/F_{\mathrm{star}}$ & & & Flux ratio between the disk and the star.\\
$f$ & &  & Scaling factor, where $0 < f < 1$.\\
$\hat{f}$ & &  & Optimum scaling factor obtained from the BFF procedure, corresponding with $\hat{D}$.\\
$H$, $H^{(k)}$, $H^{(k+1)}$ & $[H_1^{T}, \cdots, H_n^T]^T$ & $n \times N_{\text{pix}}$ & NMF component matrix for the reference cube.\\
$H_1$, $H_i$, $H_n$ &  & $ 1 \times N_{\text{pix}}$ & The $1$-st, $i$-th, and $n$-th NMF component for the reference cube ($R$).\\
${(\cdot)}^{(k)}$, ${(\cdot)}^{(k+1)}$ & superscript &  & Iteration step number.\\
$\mu_f^{(k)}$ & & & The median of the pixels in $D_f$ at iteration step $k$.\\
$N_{\text{pix}}$& &  & Number of pixels in each image.\\
$N_{\text{ref}}$& &  & Number of images in the reference cube ($R$).\\
$n$ & &  & Number of NMF components.\\
$o(\cdot)$ & &  & Little $o$ notation, meaning $|o(\cdot)| \ll |(\cdot)|$. Vectorized form means element-wise $o$'s.\\
$p_{A, B}$ & $AB^T/(BB^T)$& & The projection coefficient of row vector $A$ onto row vector $B$.\\
$R$ & $[S_1^T, S_2^T, \cdots, S_{N_{\text{ref}}}^T]^T$ & $N_{\text{ref}} \times N_{\text{pix}}$ & Reference cube with rows containing flattened references.\\
$\sigma_f^{(k)}$, $\sigma_f^{(\text{conv})}$ & & & The standard deviation for the pixels in $D_f$ at step $k$, or when the BFF procedure converges.\\
$S$, $S_i$ & & $1\times N_{\text{pix}}$ & Flattened image of only a star ($S$), subscript $i$ denotes the $i$-th star.\\
$S_{\text{NMF}}$& $\omega^{(s)} H$& $1\times N_{\text{pix}}$ & NMF model of the star only ($S$), i.e., no other astrophysical is added.\\
$(\cdot)^{T}$ & $(A^T)_{ij}=A_{ji}$ & & Transpose operator for matrices.\\
$T$ & $S + D$ & $1\times N_{\text{pix}}$ & Flattened image of a target.\\
$T_{\text{NMF}}$& $\omega H$& $1\times N_{\text{pix}}$ & NMF model of the target ($T$).\\
$V$ & & $N_{\text{ref}} \times N_{\text{pix}}$ & Variance of each pixel for the reference cube ($R$).\\
$v$ & & $1 \times N_{\text{pix}}$ & Variance of each pixel for the target image ($T$).\\
$W$, $W^{(k)}$, $W^{(k+1)}$ & & $N_{\text{ref}} \times n$ & NMF coefficient matrix for the reference cube ($R$).\\
$\omega$, $\omega^{(T)}, $$\omega^{(k)}$, $\omega^{(k+1)}$ & & $1 \times n$ & NMF coefficient matrix for the target image ($T$).\\
$\omega^{(S)}$ & & $1 \times n$ & NMF coefficient matrix for the stellar image ($S$).\\
$\omega^{({D})}$ & & $1 \times n$ & NMF coefficient matrix for the astrophysical signal ($D$).\\
$\omega_1$, $\omega_i$, $\omega_n$ & &  & The $1$-st, $i$-th, and $n$-th entry of NMF coefficient matrix for the target image ($T$).\\
\enddata 
\end{deluxetable} 

\newpage
\section{NMF with Weighting Function}\label{appendixRules}
The update rules adopted in this paper is summarized here.

1. Rules for component construction with weighting function \citep{zhu16}:
\begin{align}
W^{(k+1)} &= W^{(k)}\circ\frac{(V\circ R)H^{(k)T}}{[V\circ W^{(k)}H^{(k)}]H^{(k)T}},\\
H^{(k+1)} &= H^{(k)}\circ\frac{W^{(k)T}(V\circ R)}{W^{(k)T}[V\circ W^{(k)}H^{(k)}]},
\end{align}
where $R$ is the reference cube, $V$ is the variance matrix of the reference cube (if $V$ is not given, an empirical $V = R$ is suggested because of Poisson noise), $H^{(\cdot)}$ is the NMF component matrix for the reference cube, and $W^{(\cdot)}$ is the coefficient matrix for the reference cube. In this paper, these weighted update rules are adopted. For the {\it HST}-STIS images, the variance matrix of the reference cube is obtained from the square of the error extension in the flat-field {\tt FITS} files (i.e., the {\tt ERR} extension); when the exposures are added with simulated disks ($D$), we have $V = D + ERR^2$, where $ERR$ denotes the {\tt ERR} extension.

2. Rule for target modeling with weighting function:
\begin{equation}
\omega^{(k+1)} = \omega^{(k)}\circ\frac{(v\circ T)H^T}{[v \circ \omega^{(k)}]HH^T},
\end{equation}
where $T$ is the target, $v$ is the variance matrix of the target, $H$ is the NMF components constructed above, $\omega^{(\cdot)}$ is the coefficient matrix for the target.

\section{Construction and Stability of the Component Basis}\label{compoConstr}

As stated in the main text, we propose to use a scaling factor to correct for the capture of disk signal by NMF (see \textsection{\ref{bff}} for the detailed procedure). The use of the scaling factor is based on 1) the disk captured by NMF resembles that of the stellar PSF (\textsection{\ref{bff}}); 2) the target modeling process is linear to the first order (Appendix~\ref{scalefactorproof}); and 3) the target modeling linearity relies on the property that NMF components are stable through iteration when they are constructed {\it sequentially}, which is illustrated in this section.

There are two ways to construct components with references: set the number of components and run the iteration in Appendix~\ref{appendixRules} directly, i.e., {\it randomly}; or construct the components by starting from $1$ component, then use the corresponding coefficient and component matrices to initiate the construction for $2$ components, ..., as in \citet{zhu16}, i.e., {\it sequentially}.

When we sequentially construct the components, we can denote $w_n$ and $h_n$ as the coefficient and component matrices that are already constructed in the previous $n$ steps, and use $w_{n+1}$ and $h_{n+1}$\footnote{Note: The definition of symbols (i.e., $w_n, h_n, w_{n+1}, h_{n+1}$) here are only valid in this section for simplification (Appendix~\ref{compoConstr}), and is not included in the table of symbols in Appendix~\ref{appendixSymbols}.} as the additional coefficient and component vectors for the additional component, i.e., when $n+1$ components should be calculated. Then the simple update rules become (for clarity in this section, we only focus on the simple update rules in Eq.~\eqref{coefUpdate} \& \eqref{compUpdate}. For the weighted update rules used in this paper, the substitutions in Appendix A3 of \citealp{blanton07} should be adopted), for the coefficient matrix, 
\begin{align}
[w_n, w_{n+1}] \leftarrow [w_n, w_{n+1}] \circ \frac{R[h_n^T, h_{n+1}^T]}{[w_n, w_{n+1}] \begin{bmatrix}h_n\\ h_{n+1}\end{bmatrix} [h_n^T, h_{n+1}^T]} = [w_n, w_{n+1}] \circ \frac{[Rh_n^T, Rh_{n+1}^T]}{[w_nh_n + w_{n+1}h_{n+1}] [h_n^T, h_{n+1}^T]},
\end{align}
where the left arrow ($\leftarrow$) is a simplified notation of the updating procedure, where the left side is the result in the  $(k+1)$-th step, and the right side contains the results from the $(k)$-th step. For the component matrix,
\begin{align}
\begin{bmatrix}h_n\\ h_{n+1}\end{bmatrix} \leftarrow \begin{bmatrix}h_n\\ h_{n+1}\end{bmatrix} \circ \frac{\begin{bmatrix}w_n^T\\ w_{n+1}^T\end{bmatrix}R}{\begin{bmatrix}w_n^T\\ w_{n+1}^T\end{bmatrix} [w_n, w_{n+1}] \begin{bmatrix}h_n\\ h_{n+1}\end{bmatrix}} = \begin{bmatrix}h_n\\ h_{n+1}\end{bmatrix} \circ \frac{\begin{bmatrix}w_n^TR\\ w_{n+1}^TR\end{bmatrix}}{\begin{bmatrix}w_n^Tw_n & w_n^Tw_{n+1}\\ w_{n+1}^Tw_n & w_{n+1}^Tw_{n+1}\end{bmatrix}  \begin{bmatrix}h_n\\ h_{n+1}\end{bmatrix}}.
\end{align}

Focusing on the individual matrices, we have
\begin{align}
w_n &\leftarrow w_n \circ \frac{Rh_n^T}{w_nh_nh_n^T + w_{n+1}h_{n+1}h_n^T} = w_n \circ \frac{Rh_n^T}{w_nh_nh_n^T} \circ \frac{1}{1 + \frac{w_{n+1}h_{n+1}h_n^T}{w_nh_nh_n^T}}, \label{updateAppendixW1}\\
h_n &\leftarrow h_n \circ \frac{w_n^TR}{w_n^Tw_nh_n + w_n^Tw_{n+1}h_{n+1}} = h_n \circ \frac{w_n^TR}{w_n^Tw_nh_n} \circ \frac{1}{1 + \frac{w_n^Tw_{n+1}h_{n+1}}{w_n^Tw_nh_n}}, \label{updateAppendixH1}\\
w_{n+1} &\leftarrow w_{n+1} \circ \frac{Rh_{n+1}^T}{w_nh_nh_{n+1}^T + w_{n+1}h_{n+1}h_{n+1}^T} = w_{n+1} \circ \frac{Rh_{n+1}^T}{w_nh_nh_{n+1}^T } \circ \frac{1}{1+\frac{w_{n+1}h_{n+1}h_{n+1}^T}{w_nh_nh_{n+1}^T}}, \label{updateAppendixW2}\\
h_{n+1} &\leftarrow h_{n+1} \circ \frac{w_{n+1}^TR}{w_{n+1}^Tw_nh_n + w_{n+1}^Tw_{n+1}h_{n+1}} = h_{n+1} \circ \frac{w_{n+1}^TR}{w_{n+1}^Tw_nh_n} \circ \frac{1}{1+\frac{w_{n+1}^Tw_{n+1}h_{n+1}}{w_{n+1}^Tw_nh_n}}. \label{updateAppendixH2}
\end{align}

Given that 

({\bf a}) when $n  = m$, the coefficient matrix $w_n$ and component matrix $h_n$ already satisfy $R \approx w_n h_n$, and 

({\bf b}) the new coefficient and component vectors ($w_{n+1}$ and $h_{n+1}$) are randomly initialized \citep{zhu16}: the elements are drawn from a uniform distribution from $0$ to $1$,

the change of $h_n$ in the {\it first} iteration of in Eq.~\eqref{updateAppendixH1} before ($h_n^{\text{old}}$) and after ($h_n^{\text{new}}$) the inclusion of $w_{n+1}$ and $h_{n+1}$ gives
\begin{align}
\delta h_n &= h_n^{\text{new}} - h_n^{\text{old}}\\
		&= \left( \frac{1}{1 + \frac{w_n^Tw_{n+1}h_{n+1}}{w_n^Tw_nh_n}} - 1\right) \circ h_n \circ \frac{w_n^TR}{w_n^Tw_nh_n}\\
		&=\left( \frac{1}{1 + \frac{w_n^Tw_{n+1}h_{n+1}}{w_n^Tw_nh_n}} - 1\right) \circ h_n^{\text{old}}.\label{h1change}
\end{align}

\noindent{\bf Lemma} ({\it Stability}) : For the individual elements ($R_{(\cdot)j}$) in the reference cube, if the $R_{(\cdot)j}$'s are {\it sufficiently large} (for our purpose, they should have large signal-to-noise ratios), then the update has little impact on the constructed components (i.e., $h_n$) if we construct the components {\it sequentially}. 

\noindent{\bf Proof}: In high-contrast imaging, if the values of the pixels in the references are large, $R_{(\cdot) j}$ can be represented by 
\begin{equation}\label{eq-brightPixel}
R_{(\cdot) j} \gg 1,
\end{equation}
which, when accompanied with an weighting function as adopted in our paper (see Appendix~\ref{appendixRules}, as well as \citealp{blanton07}), the pixels should have large signal-to-noise ratios, i.e.,
\begin{equation}\label{eq-brightPixel2}
{\mathrm{SNR}}_{(\cdot) j} = \frac{R_{(\cdot) j}}{\sqrt{V_{(\cdot) j}}} \gg 1.
\end{equation}
To simplify our derivation, the above representation of signal-to-noise ratio is represented by $R_{(\cdot) j}$ in this section. This simplification is in principle valid following the substitution as in \citet{blanton07}.

Assuming that before the inclusion of the additional component, i.e., the $(m+1)$-th component (represented by $w_{n+1}$ and $h_{n+1}$), the following relationship is already satisfied during previous iterations (for $n=m$):
\begin{equation}
R_{kj} \approx (w_n h_n)_{kj},
\end{equation}
then given $w_{n+1}$ and $h_{n+1}$ are randomly initialized (drawn from a uniform distribution from 0 to 1), we have 
\begin{equation}\label{eq-faintPixel}
(w_{n+1}h_{n+1})_{kj} \le 1.
\end{equation}

Combining Eqs.~\eqref{eq-brightPixel} -- \eqref{eq-faintPixel}, we have
\begin{equation}
\frac{(w_{n+1}h_{n+1})_{kj}}{(w_n h_n)_{kj}}\ll 1,
\end{equation}
if this inequality is written in the little $o$ notation (i.e., $|o(x)| \ll |x|$), we have
\begin{equation}
\left(\frac{w_n^Tw_{n+1}h_{n+1}}{w_n^Tw_nh_n}\right)_{ij} = o(1),\label{w2h2vsw1h1}
\end{equation}
where $(w_n^Tw_nh_n)_{ij} = \sum_{k=1}^{m} (w_n)_{ik} (w_nh_n)_{kj}$ is the weighted-sum of the pixels at the same position in all the refereces.

Substitute Eq.~\eqref{w2h2vsw1h1} into Eq.~\eqref{h1change}, we have
\begin{align}
\delta (h_n)_{ij} &= \left( \frac{1}{1 + o(1)} - 1\right) (h_n)_{ij}^{\text{old}} \\
			&= \{[1+o(1)]^{-1} - 1\} (h_n)_{ij}^{\text{old}}\\
			&\approxeq -o\left((h_n)_{ij}^{\text{old}}\right),
\end{align}
and to the first order, we have equality in the above equation. In a vectorized form, we have
\begin{equation}\label{components:stable}
|\delta h_n| = |-o(h_n)| \ll |h_n|,
\end{equation}
elementwise, i.e., the addition of an extra component has little impact on the previously constructed components.
\QED

\section{Target Modeling}\label{scalefactorproof}

When the NMF components are stable through iteration, as illustrated in Appendix~\ref{compoConstr}, we are able to demonstrate the linearity of NMF modeling in this section. Assuming there are $n$ components chosen to model a target ($T$), then the $i$-th entry of the coefficient matrix $\omega$ ($i=1,\cdots,n$) in update rule Eq.~\eqref{targetmodelling} is
\begin{align}
\omega_i^{(k + 1)} &= \omega_i^{(k)}\frac{TH_i^T}{\sum_{j=1}^{n}\omega_j^{(k)}H_jH_i^T}\\
			     &= \frac{TH_i^T}{H_iH_i^T}\frac{1}{1 + \sum_{j=1, j\neq i}^n\frac{\omega_j^{(k)}}{\omega_i^{(k)}}\frac{H_jH_i^T}{H_iH_i^T}}\label{eqith}\\
			     			     &< \frac{TH_i^T}{H_iH_i^T},\label{ineq}
\end{align}
where the superscripts $^{(k)}$ and $^{(k+1)}$ are the iteration numbers, $H_i$ is the $i$-th component of $H$, and $\omega_i$ is the $i$-th entry of the coefficient matrix $\omega$. On the the right hand side of Eq.~\eqref{ineq}, it represents the coefficient of the projection of vector $T$ onto vector $H_i$.

Inequality~\eqref{ineq} arises from the fact that all the components are non-negative, therefore the denominator of the second term in Eq.~\eqref{eqith} is always larger than $1$.  This is the evidence why NMF is less prone to over-fitting -- the NMF coefficients always have smaller absolute values than normal projections: since NMF the elements are always non-negative, the normal projection coefficients are always equal to their absolute values.

When Eq.~\eqref{eqith} converges (i.e.,  $|\delta\omega_i| \ll |\omega_i|$, or $\delta\omega_i=\omega_i^{(k + 1)}-\omega_i^{(k)}=o(\omega_i)$ is satisfied), it will have a form of
\begin{equation}\label{omega_i_converge}
\omega_i = \frac{TH_i^T}{H_iH_i^T}\frac{1}{1 + \sum_{j=1, j\neq i}^n\frac{\omega_j}{\omega_i}\frac{H_jH_i^T}{H_iH_i^T}} + o(\omega_i),
\end{equation}
for simplicity, when we replace the projections with definition \begin{equation}
p_{AB} = \frac{AB^T}{BB^T},
\end{equation} Eq.~\eqref{omega_i_converge} becomes
\begin{equation}
\omega_i = p_{TH_i}\frac{1}{1 + \sum_{j=1, j\neq i}^n\frac{\omega_j}{\omega_i}p_{H_jH_i}} + o(\omega_i)= \frac{\omega_i p_{TH_i}}{\sum_{j=1}^n\omega_jp_{H_jH_i}}+ o(\omega_i).
\end{equation}

Since all the $\omega_i$'s are non-negative, dividing both sides by $\omega_i$, the above equation becomes,
\begin{equation}\label{linearity_origin}
\sum_{j=1}^n\omega_jp_{H_jH_i}=p_{TH_i}+o(p_{TH_i}).
\end{equation}

Given $T = S + D$, and using them as superscripts, we can substitute the equation into Eq.~\eqref{linearity_origin} and obtain:
\begin{equation}\begin{cases}\label{eqTSDprojections}
\sum_{j=1}^n\omega_j^{(T)}p_{H_jH_i}&=p_{TH_i}+o(p_{TH_i})\\
\sum_{j=1}^n\omega_j^{(S)}p_{H_jH_i}&=p_{SH_i}+o(p_{SH_i})\\
\sum_{j=1}^n\omega_j^{(D)}p_{H_jH_i}&=p_{DH_i}+o(p_{DH_i})\end{cases},
\end{equation}
and in addition, since
\begin{equation}
p_{TH_i} = \frac{TH_i^T}{H_iH_i^T} = \frac{(S + D)H_i^T}{H_iH_i^T} = p_{SH_i} + p_{DH_i},
\end{equation}
we have
\begin{equation}\label{TSDwithComponents}
\sum_{j=1}^n\left[\omega_j^{(T)}-(\omega_j^{(S)}+\omega_j^{(D)})\right]p_{H_jH_i} = o(p_{TH_i}).
\end{equation}

\noindent {\bf Theorem} ({\it Linearity}): The NMF target modeling process is linear to the first order when the NMF components are created sequentially and stable under iterations (i.e., when Eq.~\eqref{components:stable} in Lemma holds).

\noindent {\bf Proof}: The above equation is equivalent to proving \begin{equation}
\omega_j^{(T)}=\omega_j^{(S)}+\omega_j^{(D)} + o(\omega_j^{(T)}).\label{TSDrelation}
\end{equation}

Now we prove the above equation by way of induction:\\
A. $n = 1$, since $p_{h_nh_n} > 0$, for $i=1$, we have:
\begin{align}
\left[ \omega_1^{(T)}-(\omega_1^{(S)}+\omega_1^{(D)})\right]&p_{h_nh_n} = o(p_{Th_n}),\\
\omega_1^{(T)}-(\omega_1^{(S)}+\omega_1^{(D)})  &= o(p_{Th_n}/p_{h_nh_n}) = o(\omega_1^{(T)}),
\end{align}
Eq.~\eqref{TSDrelation} holds.\\
B. Assume for $n = m$, Eq.~\eqref{TSDrelation} holds, and we also have the following equation holds (Eq.~\eqref{TSDwithComponents}, for $i=1,\cdots,m$),
\begin{equation}
\sum_{j=1}^m\left[\omega_j^{(T)}-(\omega_j^{(S)}+\omega_j^{(D)})\right]p_{H_jH_i} = o(p_{TH_i}).\label{bwoi_b}
\end{equation}
C. For $n = m + 1$, given the fact that the components do not vary to the first order when the number of components increases (Appendix~\ref{compoConstr} Conclusion), for $i = 1, \cdots, m$, Eq.~\eqref{TSDwithComponents} becomes,
\begin{align*}
o(p_{TH_i}) 	=& \sum_{j=1}^{m+1}\left\{\omega_j^{(T)}-[\omega_j^{(S)}+\omega_j^{(D)}]\right\}p_{[H_j+o(H_j)][(H_i+o(H_i)]}\\
	=&\sum_{j=1}^m\left\{\omega_j^{(T)}-[\omega_j^{(S)}+\omega_j^{(D)}]\right\}[p_{H_jH_i} + 2o(p_{H_jH_i}) + o^2(p_{H_jH_i})] \\ 
	&+ \left\{\omega_{m+1}^{(T)}-[\omega_{m+1}^{(S)}+\omega_{m+1}^{(D)}]\right\}[p_{H_{m+1}H_i} + o(p_{H_{m+1}H_i})]\\
	=&o(p_{TH_i}) + \sum_{j=1}^m\left\{\omega_j^{(T)}-[\omega_j^{(S)}+\omega_j^{(D)}]\right\}[2o(p_{H_jH_i}) + o^2(p_{H_jH_i})]\\
	 &+ \left\{\omega_{m+1}^{(T)}-[\omega_{m+1}^{(S)}+\omega_{m+1}^{(D)}]\right\}[p_{H_{m+1}H_i} + o(p_{H_{m+1}H_i})]
\end{align*}
where Eq.~\eqref{bwoi_b} is substituted into Eq.~\eqref{TSDwithComponents} in the above derivation. Since $p_{H_{m+1}H_i}$ is a simple number rather than a vector, and Eq.~\eqref{TSDrelation} holds, by keeping up to the first order, we have
\begin{align}
\omega_{m+1}^{(T)}-[\omega_{m+1}^{(S)}+\omega_{m+1}^{(D)}] &= \frac{o(p_{TH_i})-\sum_{j=1}^m\left\{\omega_j^{(T)}-[\omega_j^{(S)}+\omega_j^{(D)}]\right\}[2o(p_{H_jH_i}) + o^2(p_{H_jH_i})]}{p_{H_{m+1}H_i} + o(p_{H_{m+1}H_i})} \\
		&= \frac{o(p_{TH_i})-2o^2(p_{TH_i}) - o^3(p_{TH_i})}{p_{H_{m+1}H_i} + o(p_{H_{m+1}H_i})} = \frac{o(p_{TH_i})}{p_{H_{m+1}H_i}}\\
		&= o(\omega_{m+1}^{(T)}),
\end{align}
which is also true when $i = m +1$, therefore the proof of Eq.~\eqref{TSDrelation} is complete.

Rewrite Eq.~\eqref{TSDrelation} in vector form, we have
\begin{equation}
\omega^{(T)} = \omega^{(S)} + \omega^{(D)} + o(\omega^{(T)}),
\end{equation}
and thus
\begin{align}
T_{\text{NMF}} &= \omega^{(T)}H = \omega^{(S)}H + \omega^{(D)}H + o(\omega^{(T)}H)\\
		     &= S_{\text{NMF}} + D_{\text{NMF}}+ o(T_{\text{NMF}}),\label{Dinfluence}
\end{align}
i.e., to the first order, we can linearly separate the stellar PSF and speckles from the circumstellar disk signal. \QED

\section{The Best Factor Finding (BFF) Procedure}\label{BFFwhy}
We notice that when the optimum scaling factor is in effect (\textsection{\ref{bff}}), the diskless regions should be well modeled by the NMF model of the target and therefore the values on these pixels should be small and have a histogram distribution that is symmetric about 0. Consequently, the variation of the noise of the diskless region should be minimized. We thus introduce the BFF procedure as follows to find this factor:

1. For each target ($T$), construct its NMF model ($T_{\text{NMF}}$) with the component basis ($H$), then vary the scaling factor $(f)$ from $0$ to $1$, creating several scaled reduced images ($D_f = T - fT_{\text{NMF}}$).

2. For each scaled reduced image ($D_f$),\\
\indent\indent (a) Identification of the background region iteratively: in each iteration ($k$), find the median ($\mu_f^{(k)}$) and standard deviation ($\sigma_f^{(k)}$) of $D_f$, remove the pixels with values satisfying the condition: 
\[{\text{value}} > \mu_f^{(k)} + 3\sigma_f^{(k)}\ \ \ \text{or}\ \ \  {\text{value}} < \mu_f^{(k)} - 10\sigma_f^{(k)}.\] 
These pixels are treated as non-background ones because of their large deviations from the median. Repeat this process until the number of background pixels does not change.\\
\indent\indent (b) Calculation of the noise in the diskless region: calculate the standard deviation of the remaining pixels when step (a) converges, and denote it by $\sigma_f^{(\text{conv})}$.

3.  The factor corresponding with the minimum standard deviation of $D_f$ of the diskless pixels will be taken as the best one ($\hat{f}$), i.e.,
\[
\hat{f} = \argmin_f \sigma_f^{(\text{conv})}.
\]

The connection between BFF and the classical optimum scaling factor is that both of them are minimizing the residual noise. In comparison, the classical method minimizes the residual noise along the major diffraction spikes after PSF subtraction \citep[e.g., ][]{schneider09}. When the diffraction spikes are not stable, especially for ground-based observations, BFF is able to focus on the entire field of view, and is more able to minimize the overall difference between the PSF template and the target.

\bibliography{refs}

\begin{thebibliography}{}
\providecommand\natexlab[1]{#1}
\providecommand\JournalTitle[1]{#1}

\bibitem[{{Amara} \& {Quanz}(2012)}]{amaya12}
{Amara}, A., \& {Quanz}, S.~P. 2012,
  \href{http://dx.doi.org/10.1111/j.1365-2966.2012.21918.x}{\JournalTitle{\mnras},
  427, 948}

\bibitem[{{Biller} {et~al.}(2004){Biller}, {Close}, {Lenzen}, {Brandner},
  {McCarthy}, {Nielsen}, \& {Hartung}}]{biller04}
{Biller}, B.~A., {Close}, L., {Lenzen}, R., {et~al.} 2004,
  \href{http://dx.doi.org/10.1117/12.552164}{in \procspie, Vol. 5490,
  Advancements in Adaptive Optics, ed. D.~{Bonaccini Calia}, B.~L.
  {Ellerbroek}, \& R.~{Ragazzoni}}, 389

\bibitem[{{Blanton} \& {Roweis}(2007)}]{blanton07}
{Blanton}, M.~R., \& {Roweis}, S. 2007,
  \href{http://dx.doi.org/10.1086/510127}{\JournalTitle{\aj}, 133, 734}

\bibitem[{{Chen} {et~al.}(2014){Chen}, {Mittal}, {Kuchner}, {Forrest}, {Lisse},
  {Manoj}, {Sargent}, \& {Watson}}]{chen14}
{Chen}, C.~H., {Mittal}, T., {Kuchner}, M., {et~al.} 2014,
  \href{http://dx.doi.org/10.1088/0067-0049/211/2/25}{\JournalTitle{\apjs},
  211, 25}

\bibitem[{{Choquet} {et~al.}(2014){Choquet}, {Pueyo}, {Hagan}, {Gofas-Salas},
  {Rajan}, {Chen}, {Perrin}, {Debes}, {Golimowski}, {Hines}, {N'Diaye},
  {Schneider}, {Mawet}, {Marois}, \& {Soummer}}]{choquet14}
{Choquet}, {\'E}., {Pueyo}, L., {Hagan}, J.~B., {et~al.} 2014,
  \href{http://dx.doi.org/10.1117/12.2056672}{in \procspie, Vol. 9143, Space
  Telescopes and Instrumentation 2014: Optical, Infrared, and Millimeter Wave},
  914357

\bibitem[{{Choquet} {et~al.}(2016){Choquet}, {Perrin}, {Chen}, {Soummer},
  {Pueyo}, {Hagan}, {Gofas-Salas}, {Rajan}, {Golimowski}, {Hines}, {Schneider},
  {Mazoyer}, {Augereau}, {Debes}, {Stark}, {Wolff}, {N'Diaye}, \&
  {Hsiao}}]{choquet16}
{Choquet}, {\'E}., {Perrin}, M.~D., {Chen}, C.~H., {et~al.} 2016,
  \href{http://dx.doi.org/10.3847/2041-8205/817/1/L2}{\JournalTitle{\apjl},
  817, L2}

\bibitem[{{Choquet} {et~al.}(2017){Choquet}, {Milli}, {Wahhaj}, {Soummer},
  {Roberge}, {Augereau}, {Booth}, {Absil}, {Boccaletti}, {Chen}, {Debes}, {del
  Burgo}, {Dent}, {Ertel}, {Girard}, {Gofas-Salas}, {Golimowski}, {G{\'o}mez
  Gonz{\'a}lez}, {Hagan}, {Hibon}, {Hines}, {Kennedy}, {Lagrange}, {Matr{\`a}},
  {Mawet}, {Mouillet}, {N'Diaye}, {Perrin}, {Pinte}, {Pueyo}, {Rajan},
  {Schneider}, {Wolff}, \& {Wyatt}}]{choquet17}
{Choquet}, {\'E}., {Milli}, J., {Wahhaj}, Z., {et~al.} 2017,
  \href{http://dx.doi.org/10.3847/2041-8213/834/2/L12}{\JournalTitle{\apjl},
  834, L12}

\bibitem[{{Debes} {et~al.}(2013){Debes}, {Jang-Condell}, {Weinberger},
  {Roberge}, \& {Schneider}}]{debes13}
{Debes}, J.~H., {Jang-Condell}, H., {Weinberger}, A.~J., {Roberge}, A., \&
  {Schneider}, G. 2013,
  \href{http://dx.doi.org/10.1088/0004-637X/771/1/45}{\JournalTitle{\apj}, 771,
  45}

\bibitem[{{Debes} {et~al.}(2017){Debes}, {Poteet}, {Jang-Condell}, {Gaspar},
  {Hines}, {Kastner}, {Pueyo}, {Rapson}, {Roberge}, {Schneider}, \&
  {Weinberger}}]{debes17}
{Debes}, J.~H., {Poteet}, C.~A., {Jang-Condell}, H., {et~al.} 2017,
  \href{http://dx.doi.org/10.3847/1538-4357/835/2/205}{\JournalTitle{\apj},
  835, 205}

\bibitem[{{Dong} \& {Fung}(2017)}]{dong17}
{Dong}, R., \& {Fung}, J. 2017,
  \href{http://dx.doi.org/10.3847/1538-4357/835/2/146}{\JournalTitle{\apj},
  835, 146}

\bibitem[{{Dong} {et~al.}(2016){Dong}, {Zhu}, {Fung}, {Rafikov}, {Chiang}, \&
  {Wagner}}]{dong16}
{Dong}, R., {Zhu}, Z., {Fung}, J., {et~al.} 2016,
  \href{http://dx.doi.org/10.3847/2041-8205/816/1/L12}{\JournalTitle{\apjl},
  816, L12}

\bibitem[{{Dong} {et~al.}(2015{\natexlab{a}}){Dong}, {Zhu}, {Rafikov}, \&
  {Stone}}]{dong15}
{Dong}, R., {Zhu}, Z., {Rafikov}, R.~R., \& {Stone}, J.~M. 2015{\natexlab{a}},
  \href{http://dx.doi.org/10.1088/2041-8205/809/1/L5}{\JournalTitle{\apjl},
  809, L5}

\bibitem[{{Dong} {et~al.}(2015{\natexlab{b}}){Dong}, {Zhu}, \&
  {Whitney}}]{dong15a}
{Dong}, R., {Zhu}, Z., \& {Whitney}, B. 2015{\natexlab{b}},
  \href{http://dx.doi.org/10.1088/0004-637X/809/1/93}{\JournalTitle{\apj}, 809,
  93}

\bibitem[{{Esposito} {et~al.}(2014){Esposito}, {Fitzgerald}, {Graham}, \&
  {Kalas}}]{esposito14}
{Esposito}, T.~M., {Fitzgerald}, M.~P., {Graham}, J.~R., \& {Kalas}, P. 2014,
  \href{http://dx.doi.org/10.1088/0004-637X/780/1/25}{\JournalTitle{\apj}, 780,
  25}

\bibitem[{{Follette} {et~al.}(2017){Follette}, {Rameau}, {Dong}, {Pueyo},
  {Close}, {Duch{\^e}ne}, {Fung}, {Leonard}, {Macintosh}, {Males}, {Marois},
  {Millar-Blanchaer}, {Morzinski}, {Mullen}, {Perrin}, {Spiro}, {Wang},
  {Ammons}, {Bailey}, {Barman}, {Bulger}, {Chilcote}, {Cotten}, {De Rosa},
  {Doyon}, {Fitzgerald}, {Goodsell}, {Graham}, {Greenbaum}, {Hibon}, {Hung},
  {Ingraham}, {Kalas}, {Konopacky}, {Larkin}, {Maire}, {Marchis}, {Metchev},
  {Nielsen}, {Oppenheimer}, {Palmer}, {Patience}, {Poyneer}, {Rajan},
  {Rantakyr{\"o}}, {Savransky}, {Schneider}, {Sivaramakrishnan}, {Song},
  {Soummer}, {Thomas}, {Vega}, {Wallace}, {Ward-Duong}, {Wiktorowicz}, \&
  {Wolff}}]{follette17}
{Follette}, K.~B., {Rameau}, J., {Dong}, R., {et~al.} 2017,
  \href{http://dx.doi.org/10.3847/1538-3881/aa6d85}{\JournalTitle{\aj}, 153,
  264}

\bibitem[{{Gomez Gonzalez} {et~al.}(2017){Gomez Gonzalez}, {Wertz}, {Absil},
  {Christiaens}, {Defr{\`e}re}, {Mawet}, {Milli}, {Absil}, {Van Droogenbroeck},
  {Cantalloube}, {Hinz}, {Skemer}, {Karlsson}, \& {Surdej}}]{gonzalez17}
{Gomez Gonzalez}, C.~A., {Wertz}, O., {Absil}, O., {et~al.} 2017,
  \href{http://dx.doi.org/10.3847/1538-3881/aa73d7}{\JournalTitle{\aj}, 154, 7}

\bibitem[{{Grady} {et~al.}(2010){Grady}, {Hamaguchi}, {Schneider}, {Stecklum},
  {Woodgate}, {McCleary}, {Williger}, {Sitko}, {M{\'e}nard}, {Henning},
  {Brittain}, {Troutmann}, {Donehew}, {Hines}, {Wisniewski}, {Lynch},
  {Russell}, {Rudy}, {Day}, {Shenoy}, {Wilner}, {Silverstone}, {Bouret},
  {Meusinger}, {Clampin}, {Kim}, {Petre}, {Sahu}, {Endres}, \&
  {Collins}}]{grady10}
{Grady}, C.~A., {Hamaguchi}, K., {Schneider}, G., {et~al.} 2010,
  \href{http://dx.doi.org/10.1088/0004-637X/719/2/1565}{\JournalTitle{\apj},
  719, 1565}

\bibitem[{{Grady} {et~al.}(2013){Grady}, {Muto}, {Hashimoto}, {Fukagawa},
  {Currie}, {Biller}, {Thalmann}, {Sitko}, {Russell}, {Wisniewski}, {Dong},
  {Kwon}, {Sai}, {Hornbeck}, {Schneider}, {Hines}, {Moro Mart{\'{\i}}n},
  {Feldt}, {Henning}, {Pott}, {Bonnefoy}, {Bouwman}, {Lacour}, {Mueller},
  {Juh{\'a}sz}, {Crida}, {Chauvin}, {Andrews}, {Wilner}, {Kraus}, {Dahm},
  {Robitaille}, {Jang-Condell}, {Abe}, {Akiyama}, {Brandner}, {Brandt},
  {Carson}, {Egner}, {Follette}, {Goto}, {Guyon}, {Hayano}, {Hayashi},
  {Hayashi}, {Hodapp}, {Ishii}, {Iye}, {Janson}, {Kandori}, {Knapp}, {Kudo},
  {Kusakabe}, {Kuzuhara}, {Mayama}, {McElwain}, {Matsuo}, {Miyama}, {Morino},
  {Nishimura}, {Pyo}, {Serabyn}, {Suto}, {Suzuki}, {Takami}, {Takato},
  {Terada}, {Tomono}, {Turner}, {Watanabe}, {Yamada}, {Takami}, {Usuda}, \&
  {Tamura}}]{grady13}
{Grady}, C.~A., {Muto}, T., {Hashimoto}, J., {et~al.} 2013,
  \href{http://dx.doi.org/10.1088/0004-637X/762/1/48}{\JournalTitle{\apj}, 762,
  48}

\bibitem[{{Hinkley} {et~al.}(2007){Hinkley}, {Oppenheimer}, {Soummer},
  {Sivaramakrishnan}, {Roberts}, {Kuhn}, {Makidon}, {Perrin}, {Lloyd},
  {Kratter}, \& {Brenner}}]{hinkley07}
{Hinkley}, S., {Oppenheimer}, B.~R., {Soummer}, R., {et~al.} 2007,
  \href{http://dx.doi.org/10.1086/509063}{\JournalTitle{\apj}, 654, 633}

\bibitem[{{Jang-Condell} \& {Boss}(2007)}]{jangcondell07}
{Jang-Condell}, H., \& {Boss}, A.~P. 2007,
  \href{http://dx.doi.org/10.1086/516837}{\JournalTitle{\apjl}, 659, L169}

\bibitem[{{Jang-Condell} \& {Turner}(2012)}]{jangcondell12}
{Jang-Condell}, H., \& {Turner}, N.~J. 2012,
  \href{http://dx.doi.org/10.1088/0004-637X/749/2/153}{\JournalTitle{\apj},
  749, 153}

\bibitem[{{Jang-Condell} \& {Turner}(2013)}]{jangcondell13}
---. 2013,
  \href{http://dx.doi.org/10.1088/0004-637X/772/1/34}{\JournalTitle{\apj}, 772,
  34}

\bibitem[{{Krist} {et~al.}(2011){Krist}, {Hook}, \& {Stoehr}}]{krist11}
{Krist}, J.~E., {Hook}, R.~N., \& {Stoehr}, F. 2011,
  \href{http://dx.doi.org/10.1117/12.892762}{in \procspie, Vol. 8127, Optical
  Modeling and Performance Predictions V}, 81270J

\bibitem[{{Lafreni{\`e}re} {et~al.}(2009){Lafreni{\`e}re}, {Marois}, {Doyon},
  \& {Barman}}]{lafreniere09}
{Lafreni{\`e}re}, D., {Marois}, C., {Doyon}, R., \& {Barman}, T. 2009,
  \href{http://dx.doi.org/10.1088/0004-637X/694/2/L148}{\JournalTitle{\apjl},
  694, L148}

\bibitem[{{Lafreni{\`e}re} {et~al.}(2007){Lafreni{\`e}re}, {Marois}, {Doyon},
  {Nadeau}, \& {Artigau}}]{lafreniere07}
{Lafreni{\`e}re}, D., {Marois}, C., {Doyon}, R., {Nadeau}, D., \& {Artigau},
  {\'E}. 2007, \href{http://dx.doi.org/10.1086/513180}{\JournalTitle{\apj},
  660, 770}

\bibitem[{Lee \& Seung(2001)}]{lee01}
Lee, D.~D., \& Seung, H.~S. 2001,
  \href{http://papers.nips.cc/paper/1861-algorithms-for-non-negative-matrix-factorization.pdf}{in
  Advances in Neural Information Processing Systems 13, ed. T.~K. Leen, T.~G.
  Dietterich, \& V.~Tresp} (MIT Press), 556

\bibitem[{{Lee} \& {Chiang}(2016)}]{lee16}
{Lee}, E.~J., \& {Chiang}, E. 2016,
  \href{http://dx.doi.org/10.3847/0004-637X/827/2/125}{\JournalTitle{\apj},
  827, 125}

\bibitem[{{Marino} {et~al.}(2016){Marino}, {Matr{\`a}}, {Stark}, {Wyatt},
  {Casassus}, {Kennedy}, {Rodriguez}, {Zuckerman}, {Perez}, {Dent}, {Kuchner},
  {Hughes}, {Schneider}, {Steele}, {Roberge}, {Donaldson}, \&
  {Nesvold}}]{marino16}
{Marino}, S., {Matr{\`a}}, L., {Stark}, C., {et~al.} 2016,
  \href{http://dx.doi.org/10.1093/mnras/stw1216}{\JournalTitle{\mnras}, 460,
  2933}

\bibitem[{{Marois} {et~al.}(2006){Marois}, {Lafreni{\`e}re}, {Doyon},
  {Macintosh}, \& {Nadeau}}]{marois06}
{Marois}, C., {Lafreni{\`e}re}, D., {Doyon}, R., {Macintosh}, B., \& {Nadeau},
  D. 2006, \href{http://dx.doi.org/10.1086/500401}{\JournalTitle{\apj}, 641,
  556}

\bibitem[{{Marois} {et~al.}(2010){Marois}, {Macintosh}, \&
  {V{\'e}ran}}]{marois10}
{Marois}, C., {Macintosh}, B., \& {V{\'e}ran}, J.-P. 2010,
  \href{http://dx.doi.org/10.1117/12.857225}{in \procspie, Vol. 7736, Adaptive
  Optics Systems II}, 77361J

\bibitem[{{Mawet} {et~al.}(2012){Mawet}, {Pueyo}, {Lawson}, {Mugnier}, {Traub},
  {Boccaletti}, {Trauger}, {Gladysz}, {Serabyn}, {Milli}, {Belikov}, {Kasper},
  {Baudoz}, {Macintosh}, {Marois}, {Oppenheimer}, {Barrett}, {Beuzit},
  {Devaney}, {Girard}, {Guyon}, {Krist}, {Mennesson}, {Mouillet}, {Murakami},
  {Poyneer}, {Savransky}, {V{\'e}rinaud}, \& {Wallace}}]{mawet12}
{Mawet}, D., {Pueyo}, L., {Lawson}, P., {et~al.} 2012,
  \href{http://dx.doi.org/10.1117/12.927245}{in \procspie, Vol. 8442, Space
  Telescopes and Instrumentation 2012: Optical, Infrared, and Millimeter Wave},
  844204

\bibitem[{{Mazoyer} {et~al.}(2014){Mazoyer}, {Boccaletti}, {Augereau},
  {Lagrange}, {Galicher}, \& {Baudoz}}]{mazoyer14}
{Mazoyer}, J., {Boccaletti}, A., {Augereau}, J.-C., {et~al.} 2014,
  \href{http://dx.doi.org/10.1051/0004-6361/201424479}{\JournalTitle{\aap},
  569, A29}

\bibitem[{{Mazoyer} {et~al.}(2016){Mazoyer}, {Boccaletti}, {Choquet}, {Perrin},
  {Pueyo}, {Augereau}, {Lagrange}, {Debes}, \& {Wolff}}]{mazoyer16}
{Mazoyer}, J., {Boccaletti}, A., {Choquet}, {\'E}., {et~al.} 2016,
  \href{http://dx.doi.org/10.3847/0004-637X/818/2/150}{\JournalTitle{\apj},
  818, 150}

\bibitem[{{Milli} {et~al.}(2014){Milli}, {Lagrange}, {Mawet}, {Absil},
  {Augereau}, {Mouillet}, {Boccaletti}, {Girard}, \& {Chauvin}}]{milli14}
{Milli}, J., {Lagrange}, A.-M., {Mawet}, D., {et~al.} 2014,
  \href{http://dx.doi.org/10.1051/0004-6361/201323130}{\JournalTitle{\aap},
  566, A91}

\bibitem[{{Nesvold} {et~al.}(2016){Nesvold}, {Naoz}, {Vican}, \&
  {Farr}}]{nesvold16}
{Nesvold}, E.~R., {Naoz}, S., {Vican}, L., \& {Farr}, W.~M. 2016,
  \href{http://dx.doi.org/10.3847/0004-637X/826/1/19}{\JournalTitle{\apj}, 826,
  19}

\bibitem[{{Paatero} \& {Tapper}(1994)}]{paatero94}
{Paatero}, P., \& {Tapper}, U. 1994,
  \href{http://dx.doi.org/10.1002/env.3170050203}{\JournalTitle{Environmetrics},
  5, 111}

\bibitem[{{Perrin} {et~al.}(2004){Perrin}, {Graham}, {Kalas}, {Lloyd}, {Max},
  {Gavel}, {Pennington}, \& {Gates}}]{perrin04}
{Perrin}, M.~D., {Graham}, J.~R., {Kalas}, P., {et~al.} 2004,
  \href{http://dx.doi.org/10.1126/science.1094602}{\JournalTitle{Science}, 303,
  1345}

\bibitem[{{Perrin} {et~al.}(2008){Perrin}, {Graham}, \& {Lloyd}}]{perrin08}
{Perrin}, M.~D., {Graham}, J.~R., \& {Lloyd}, J.~P. 2008,
  \href{http://dx.doi.org/10.1086/588153}{\JournalTitle{\pasp}, 120, 555}

\bibitem[{{Pinte} {et~al.}(2009){Pinte}, {Harries}, {Min}, {Watson},
  {Dullemond}, {Woitke}, {M{\'e}nard}, \& {Dur{\'a}n-Rojas}}]{pinte09}
{Pinte}, C., {Harries}, T.~J., {Min}, M., {et~al.} 2009,
  \href{http://dx.doi.org/10.1051/0004-6361/200811555}{\JournalTitle{\aap},
  498, 967}

\bibitem[{{Pinte} {et~al.}(2006){Pinte}, {M{\'e}nard}, {Duch{\^e}ne}, \&
  {Bastien}}]{pinte06}
{Pinte}, C., {M{\'e}nard}, F., {Duch{\^e}ne}, G., \& {Bastien}, P. 2006,
  \href{http://dx.doi.org/10.1051/0004-6361:20053275}{\JournalTitle{\aap}, 459,
  797}

\bibitem[{{Pueyo}(2016)}]{pueyo16}
{Pueyo}, L. 2016,
  \href{http://dx.doi.org/10.3847/0004-637X/824/2/117}{\JournalTitle{\apj},
  824, 117}

\bibitem[{{Pueyo} {et~al.}(2012){Pueyo}, {Crepp}, {Vasisht}, {Brenner},
  {Oppenheimer}, {Zimmerman}, {Hinkley}, {Parry}, {Beichman}, {Hillenbrand},
  {Roberts}, {Dekany}, {Shao}, {Burruss}, {Bouchez}, {Roberts}, \&
  {Soummer}}]{pueyo12}
{Pueyo}, L., {Crepp}, J.~R., {Vasisht}, G., {et~al.} 2012,
  \href{http://dx.doi.org/10.1088/0067-0049/199/1/6}{\JournalTitle{\apjs}, 199,
  6}

\bibitem[{{Pueyo} {et~al.}(2015){Pueyo}, {Soummer}, {Hoffmann}, {Oppenheimer},
  {Graham}, {Zimmerman}, {Zhai}, {Wallace}, {Vescelus}, {Veicht}, {Vasisht},
  {Truong}, {Sivaramakrishnan}, {Shao}, {Roberts}, {Roberts}, {Rice}, {Parry},
  {Nilsson}, {Lockhart}, {Ligon}, {King}, {Hinkley}, {Hillenbrand}, {Hale},
  {Dekany}, {Crepp}, {Cady}, {Burruss}, {Brenner}, {Beichman}, \&
  {Baranec}}]{pueyo15}
{Pueyo}, L., {Soummer}, R., {Hoffmann}, J., {et~al.} 2015,
  \href{http://dx.doi.org/10.1088/0004-637X/803/1/31}{\JournalTitle{\apj}, 803,
  31}

\bibitem[{{Ren} {et~al.}(2017){Ren}, {Pueyo}, {Perrin}, {Debes}, \&
  {Choquet}}]{ren17}
{Ren}, B., {Pueyo}, L., {Perrin}, M.~D., {Debes}, J.~H., \& {Choquet}, E. 2017,
  \href{http://dx.doi.org/10.1117/12.2274163}{in \procspie, Vol. 10400,
  Techniques and Instrumentation for Detection of Exoplanets VIII}, 1040021

\bibitem[{{Riley}(2017)}]{stisihb17}
{Riley}, A. 2017, {STIS Instrument Hanbook for Cycle 25, Version 16.0}

\bibitem[{{Rodigas} {et~al.}(2014){Rodigas}, {Malhotra}, \& {Hinz}}]{rodigas14}
{Rodigas}, T.~J., {Malhotra}, R., \& {Hinz}, P.~M. 2014,
  \href{http://dx.doi.org/10.1088/0004-637X/780/1/65}{\JournalTitle{\apj}, 780,
  65}

\bibitem[{{Schneider} {et~al.}(2009){Schneider}, {Weinberger}, {Becklin},
  {Debes}, \& {Smith}}]{schneider09}
{Schneider}, G., {Weinberger}, A.~J., {Becklin}, E.~E., {Debes}, J.~H., \&
  {Smith}, B.~A. 2009,
  \href{http://dx.doi.org/10.1088/0004-6256/137/1/53}{\JournalTitle{\aj}, 137,
  53}

\bibitem[{{Schneider} {et~al.}(2014){Schneider}, {Grady}, {Hines}, {Stark},
  {Debes}, {Carson}, {Kuchner}, {Perrin}, {Weinberger}, {Wisniewski},
  {Silverstone}, {Jang-Condell}, {Henning}, {Woodgate}, {Serabyn},
  {Moro-Martin}, {Tamura}, {Hinz}, \& {Rodigas}}]{schneider14}
{Schneider}, G., {Grady}, C.~A., {Hines}, D.~C., {et~al.} 2014,
  \href{http://dx.doi.org/10.1088/0004-6256/148/4/59}{\JournalTitle{\aj}, 148,
  59}

\bibitem[{{Schneider} {et~al.}(2016){Schneider}, {Grady}, {Stark}, {Gaspar},
  {Carson}, {Debes}, {Henning}, {Hines}, {Jang-Condell}, {Kuchner}, {Perrin},
  {Rodigas}, {Tamura}, \& {Wisniewski}}]{schneider16}
{Schneider}, G., {Grady}, C.~A., {Stark}, C.~C., {et~al.} 2016,
  \href{http://dx.doi.org/10.3847/0004-6256/152/3/64}{\JournalTitle{\aj}, 152,
  64}

\bibitem[{{Smith} \& {Terrile}(1984)}]{smith84}
{Smith}, B.~A., \& {Terrile}, R.~J. 1984,
  \href{http://dx.doi.org/10.1126/science.226.4681.1421}{\JournalTitle{Science},
  226, 1421}

\bibitem[{{Soummer} {et~al.}(2007){Soummer}, {Ferrari}, {Aime}, \&
  {Jolissaint}}]{soummer07}
{Soummer}, R., {Ferrari}, A., {Aime}, C., \& {Jolissaint}, L. 2007,
  \href{http://dx.doi.org/10.1086/520913}{\JournalTitle{\apj}, 669, 642}

\bibitem[{{Soummer} {et~al.}(2012){Soummer}, {Pueyo}, \& {Larkin}}]{soummer12}
{Soummer}, R., {Pueyo}, L., \& {Larkin}, J. 2012,
  \href{http://dx.doi.org/10.1088/2041-8205/755/2/L28}{\JournalTitle{\apjl},
  755, L28}

\bibitem[{{Soummer} {et~al.}(2014){Soummer}, {Perrin}, {Pueyo}, {Choquet},
  {Chen}, {Golimowski}, {Brendan Hagan}, {Mittal}, {Moerchen}, {N'Diaye},
  {Rajan}, {Wolff}, {Debes}, {Hines}, \& {Schneider}}]{soummer14}
{Soummer}, R., {Perrin}, M.~D., {Pueyo}, L., {et~al.} 2014,
  \href{http://dx.doi.org/10.1088/2041-8205/786/2/L23}{\JournalTitle{\apjl},
  786, L23}

\bibitem[{{Stark} {et~al.}(2014){Stark}, {Schneider}, {Weinberger}, {Debes},
  {Grady}, {Jang-Condell}, \& {Kuchner}}]{stark14}
{Stark}, C.~C., {Schneider}, G., {Weinberger}, A.~J., {et~al.} 2014,
  \href{http://dx.doi.org/10.1088/0004-637X/789/1/58}{\JournalTitle{\apj}, 789,
  58}

\bibitem[{{Traub} \& {Oppenheimer}(2010)}]{traub10}
{Traub}, W.~A., \& {Oppenheimer}, B.~R. 2010, {Direct Imaging of Exoplanets},
  ed. S.~{Seager}, 111

\bibitem[{{Wahhaj} {et~al.}(2015){Wahhaj}, {Cieza}, {Mawet}, {Yang}, {Canovas},
  {de Boer}, {Casassus}, {M{\'e}nard}, {Schreiber}, {Liu}, {Biller}, {Nielsen},
  \& {Hayward}}]{wahhaj15}
{Wahhaj}, Z., {Cieza}, L.~A., {Mawet}, D., {et~al.} 2015,
  \href{http://dx.doi.org/10.1051/0004-6361/201525837}{\JournalTitle{\aap},
  581, A24}

\bibitem[{{Wang} {et~al.}(2015){Wang}, {Ruffio}, {De Rosa}, {Aguilar}, {Wolff},
  \& {Pueyo}}]{pyklip15}
{Wang}, J.~J., {Ruffio}, J.-B., {De Rosa}, R.~J., {et~al.} 2015, {pyKLIP: PSF
  Subtraction for Exoplanets and Disks}, Astrophysics Source Code Library,
  \href{http://arxiv.org/abs/1506.001}{{\sffamily ascl:1506.001}}

\bibitem[{{Zhu}(2016)}]{zhu16}
{Zhu}, G. 2016, \JournalTitle{ArXiv e-prints},
  \href{http://arxiv.org/abs/1612.06037}{{\sffamily arXiv:1612.06037
  [astro-ph.IM]}}

\bibitem[{{Zhu} {et~al.}(2015){Zhu}, {Dong}, {Stone}, \& {Rafikov}}]{zhu15}
{Zhu}, Z., {Dong}, R., {Stone}, J.~M., \& {Rafikov}, R.~R. 2015,
  \href{http://dx.doi.org/10.1088/0004-637X/813/2/88}{\JournalTitle{\apj}, 813,
  88}

\end{thebibliography}

\end{CJK*}
\end{document}